# Biological effects of low power nonionizing radiation: A Narrative Review

Biswadev Roy[1*], Suryakant Niture[2], and Marvin H. Wu[1]

[1] Department of Mathematics & Physics, North Carolina Central University, 1900 Concord Street, N.C. 27707, U.S.A.

[2] Julius L. Chambers Biomedical/Biotechnology Research Institute (BBRI), North Carolina Central University, Durham, NC 27707, U.S.A.

**Abstract**

Background and controlled electromagnetic radiation (EMR) on cells and tissues induces thermal, non-thermal, and dielectric property change. After EMR interaction with cells/tissues the resulting signal is used for imaging, bio-molecular response, and photo-biomodulation studies at infrared regime, and therapeutic use. We attempt to present a review of current literature with a focus to present compilation of published experimental results for each regime viz. microwave (extremely low frequency, ELF to 3 GHz), to cellular communication frequencies (100 kHz to 300 GHz), millimeter wave (300 GHz- 1 THz), THz (1 THz-20THz) and the infra-red band extending up to 461 THz. A graphical representation of the frequency effects and their relevant significance in detection of direct biological effects, therapeutic applications and biophysical interpretation is presented. A total of seventy research papers from peer-reviewed journals were used to compile a mixture of useful information, all presented in a narrative style. Out of the Journal articles used for this paper, 63 journal articles were published between 2000 to 2020. Physical, biological, and therapeutic mechanisms of thermal, non-thermal and complex dielectric effects of EMR on cells are all explained in relevant sections of this paper. A broad up to date review for the EMR range kHz-NIR (kilohertz to near infra-red) is prepared. Published reports indicate that number of biological cell irradiation impact studies fall off rapidly beyond a few THz EMR, leading to relatively a smaller number of studies in FIR and NIR bands covering most of the thermal effects and microthermal effects, and rotation-vibration effects.

[*] Corresponding author:  Biswadev Roy, Department of Mathematics & Physics, North Carolina Central University, #1146, Mary Townes Science Building, 1900 Concord Street, Durham, N.C.  27560, email:  broy@nccu.edu

# 1. Introduction

Human cells and tissues are continuously exposed to background and man-made electromagnetic radiation (EMR) emanating from a variety of sources. The EMR interacts with cells by imposing the electric field and magnetic field to cause charges inside the cell to redistribute, giving rise to a relaxation period that obey the laws of electromagnetism and thermodynamics. EMR (alternating field) impulse on the biological cell either result in dipole relaxation, electron or atomic or ionic polarization. Quite a bit of data has already been acquired in many different areas of biological sciences in which EMR is either used for spectroscopic studies or used as a carrier signal that can decipher the underlying changes in dielectric properties of the illuminated cell. These published papers with data are peer-reviewed and are likely to be credible to the scientific community. Each experiment cited in this paper is carefully designed, and authors have clearly described how the data is quality controlled.

In this review we present the state of the science about experimental evidence of biological effects resulting from non-ionizing radiation exposures at the cellular level at heavily assigned radio- to near infra-red frequencies, which are widely used for communication, surveillance, controls, and medical diagnostic equipment or in hospitals. Although, biological organisms are electrically neutral, and EMR supports homeostasis, because the constituents are being subatomic/charged particles, visible light tends to be absorbed directly by electrons in the biomaterial, whereas the higher frequency EMRs, such as terahertz (THz) and infrared (IR) energy, tend to be absorbed by the bonds within molecules. This results in a natural response that cause enhancement of vibration, bond stretching, twisting, bending, etc. There is a probability of thermal conversion also within the tissue. We attempt to illustrate, elucidate, categorize source and receptor consequences in terms of biophysical/biochemical, radiation safety and medical manifests, as well as discuss a broad experimental validation work presented in the current literature. There are a variety of biological effects resulting from exposure of tissues to different frequency regimes from radio- to the IR domains. Broadly, from radio-communication point of view we can consider the nomenclature of bands as extremely low frequency (ELF) (3Hz-3kHz), 0.3 GHz to 3 GHz as "radiofrequency (RF), or "radio" domain, 3 GHz to 30 GHz as "Microwave" domain, 30 GHz-300 GHz as "millimeter wave (mmW)", 300 GHz to 10 x 20$^{12}$ Hz as "Terahertz" and 20 THz-461 THz as the "infrared (IR)" domain . Sometimes we may term any frequency electromagnetic radiation (EMR) propagation to be an "RF" signal.

The areas of discussion in this review are based on exposure to EMR sources that have characteristic frequencies in the range ELF-300GHz mainly due to RF emission by RF electrical communication systems, medical, and industrial processes. As seen from Figure 1, the electromagnetic (EM) spectrum shows the frequency and sizes of the radio waves in comparison to the size of the mundane objects. It needs to be pointed out that the artificial emission of radio-wave that affect all biological systems is only a small fraction of the entire spectrum. Figure 1 also shows the EMR spectrum with wavelengths and wavenumbers compared by size, corresponding energy levels of photons, and the temperature of radiation.

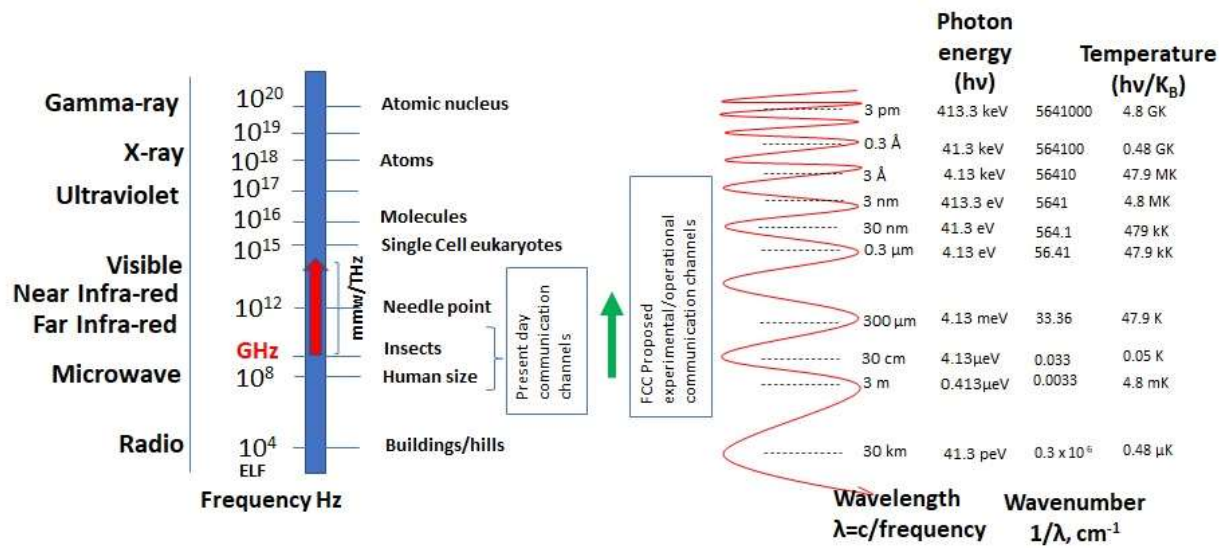

**Figure 1.** The entire EMR spectrum (ELF to $10^{20}$ Hz) with highlight of the present-day RF communication channels, and projected frequency domain based on FCC proposal for future communication links

Mobile telephony is the most ubiquitous source of artificial RF energy in today's environment. All RF communication activity at present is taking place in the 0.8 GHz - 47 GHz (FCC 2018a). The Federal Communications Commission (FCC) has opened use of frequencies extending up to 3000 GHz (FCC 2018b) to perform experiments and testing applicability. With the current and future RF bands operating (100 kHz-60 GHz), we consider biological cell exposures to photon energies in electron-volts between 0.41 neV (nano electron-volt) to about 248 μeV (micro electron-volt), and corresponding blackbody temperature 5μK to about 2.9K, which is a considerable temperature rise in the far end of the permissible band. In these communication RF bands, the photon energies are not enough to ionize or remove the valence electrons from the water and biomolecules, hence, these are termed "non-ionizing" radiation. This prevents the creation of free-radicals, or secondary damage to the biological molecules. Wilmink and Grundt (2011), and Romanenko et al. (2017) published a very informative review titled, "Current state of research on biological effects of terahertz radiation" in which THz spectral band is explained, followed by compositional information about the skin-tissue, mammalian cell, organelles, and biological macromolecules. There is almost a 6-year lag between the comprehensive review work done by Wilmink and Grundt (2011) and those by Romanenko et al. (2017). This latter reference also presented some results on the work carried out in theoretical aspects, and to those which are related to experimental outcomes. It is noted that many authors have presented the biological effects of the incident millimeter wave power versus mmW energy. Many can argue that due to presence of water, power density and resulting thermal effect in biological tissues, all become a very important parameter of the stimulating EMR. However, it is also pointed out that power density can be compensated by exposure-time to accomplish physiological change.

EMR illumination of cells and tissues have two-fold implications that are direct. One being the thermal effects on non-ionizing radiation and the other being the non-thermal effects. These effects have been observed almost in the entire radio spectrum extending from extremely low frequency (ELF)-RF-microwave-mmW-THz and up to the near infra-red (NIR) band. In terms of exposure, a detailed study was reported by the National Academies where descriptions of *in-vivo* studies about pulsed and continuous wave (CW) EMR, have been given (The National Academies 1993). In the review article mentioned earlier (Wilmink and Grundt 2011), complete biophysical aspects of EMR exposure (in vitro, ex vivo and in vivo) from assessment of the thermal, and microthermal effects have been presented. A critical discussion of the 2001-2004 project "THz-BRIDGE" (THz radiation in Biological Research, Investigation on Diagnostics, and study on potential Genotoxic Effects) along with slight description of THz-BEAM project (THz radiation in Biological, Environmental and Advanced Materials studies) have been presented. This reference carries a detailed summary of the cellular effects associated with THz radiation (0.1 to 3.7 THz). Authors have carefully pointed out that enough details are required for exposure-duration, type of source used, detailed specifications of detectors, and about the temperature-controlled exposure chamber for *in vitro* studies as well as the use of empirical and computational modeling.

Thermal transients are developed due to optical energy being converted into heat within the biological materials. In the absence of photochemical activity and phase transitions, all energy absorbed gives rise to increase in tissue temperature and thermal diffusion in 3 dimensions can be derived. The rate of heat generation (Wilmink and Grundt, 2011) in Watts/m$^3$

$$S(r,z) = k_a(r,z)\Phi_0(r,z) \qquad (1)$$

Where S is the heat generation rate (W/m$^3$), $k_a$ is the local absorption coefficient at point (r,z) and $\Phi_0(r,z)$ is the irradiance at the point. Using S and tissue heat capacity for $\Delta t$ duration of heat exposure we can compute the local rise in temperature easily:

$$\Delta T(r,z) = \frac{S(r,z)\Delta t}{\rho c} \qquad (2)$$

ρ is the tissue density and c are the tissue's specific heat capacity in Joule/gram/Kelvin. Using temperature rise ΔT the heat transfer can be modeled (Penne's bioheat equation)

$$\rho c \frac{\partial T}{\partial t} = K\nabla^2 T + S + \bar{q} \qquad (3)$$

K is the thermal conductivity (Watts/m/K), ρ is tissue/water density in kg/m$^3$ T is the temperature in Kelvin, S is the heat generation rate (Eq. 1), and q is the perfusion (heat transfer) rate. Eq. 3 can be described in detail (Dehgan and Sabouri, 2012) as

$$\rho c \frac{\partial T(r,t)}{\partial t} = \nabla \cdot [K\nabla T(r,t)] + w_b \rho_b c_b [T_a - T(r,t)] + Q_m + Q_r \qquad (4)$$

T is the tissue temperature at moment t, having position vector $r = [r_1, r_2, \ldots r_d]$, d being the spatial dimension $\rho_b$, and $c_b$ are density and specific heat of blood, $w_b$ is blood perfusion rate, $Q_m$ and $Q_r$ are volumetric heat sources due to metabolism and spatial heating respectively and $T_a$ is the arterial temperature. Equation 4 can only be solved using numerical methods. Amongst the numerical methods such as finite element, finite difference, and boundary element methods, a spectral method yields a high accuracy solution(Dehgan and Sabouri, 2012)

Shapiro et al. (2013) have studied thermal mechanisms in detail for the mmW spectrum (30 to 300 GHz). In mmW domain, thermal mechanisms are more pronounced since the absorbance to water is remarkably large. At these frequencies optical energy is directly absorbed by the target chromophores, and as a result we get thermal transients that can be recorded. Neglecting photochemical effects and phase transitions, it is possible to compute the magnitude and rate of energy deposition on the tissue, for example, by knowing the skin absorption rate. Thus, it makes it possible to model the 3-dimensional propagation of THz photons inside bio-tissue, which is of a great advantage. Thermal effects on mammalian cells have also been reported in detail by Wilmink and Grundt (2011). These effects include stimulation of cell growth and metabolism, morphological changes (swelling, etc.) activation of cellular stress response (CSR) mechanisms (Wilmink et al. 2006, 2008, 2009, 2010), as well as apoptopic and necrotic pathways. Time-temperature dosimetry, and region-based analysis is suggested by the authors. Mild thermal stress (40-42$^0$C rise for exposure period 10 to 30 minutes) has been shown to increase growth and metabolic rates by 20% (Wilmink et al. 2008, 2009; Bethuan et al. 2004; F.M. -P de Gannes et al. 1998). It is also reported that temperature ranging between 40-42$^0$C is not lethal to cells but can lead to alterations to morphology such as cell flattening, membrane ruffling, etc. Mild hyperthermia due to THz exposure leads to activation of the intracellular signaling pathways.

Mammalian cells exposed to temperatures that range between 42-46$^0$C for 30-50 minutes, exhibit dramatically altered morphology, and inhibit cell cycle progression and activation of CSR (a molecular defense mechanism). Mammalian cells exposed to temperatures greater that 46$^0$C lead to cell shape collapse, rupture of plasma and nuclear membrane and ultimately cell death. Romanenko et al. (2017) provided many references that spell out the nonlinear biological phenomena and some theoretical basis for such claims made. These authors have also pointed out that some types of electric and magnetic fields act as cell homeostatic agents and provide many relevant references therein that reveal details. Notably, authors point out that organisms contain electrically charged, neutral, polarized molecules, and the resultant electric fields obey all laws of electromagnetism and thermodynamics. A dead cell will have zero transmembrane potential, and the existence of an electrical potential across cell membrane will create chemical gradients between intra- and extracellular spaces. In experiments with neurons and neuroblasts, authors mention that neurons and neuroblasts of vertebrate, as well as invertebrate species external electric field, causes preferential growth of neurites that have cathode affinity (attracted towards the negative electrode). Biological objects interact with static and alternating EMR fields, with this being said, the degree of biological interaction depends upon the type of organism, tissue, molecular composition of the cell, and on the EMR parameters such as the frequency, power density, modulation, polarization, pulse mode, power and total energy absorbed.

Non-thermal effects in cells (Alexandrov et al. 2009, 2010; Chitnavis 2006; Reimers et al. 2009) are also able to be studied using microwave/THz beam. This phenomenon has resemblance to the interaction of microwave or mmW to photocarriers inside the solar materials. THz has oscillations with about $10^{12}$ cycles per second which match up with the natural phonon frequencies of the bio-sample (Chitnavis 2006; Alexandrov et al. 2010). THz interactions with DNA molecules may disrupt the hydrogen bond between DNA strands resulting in structural alteration in double stranded DNA molecule. Authors mentioned that no experimental details were gathered in this line because of the detection issues, especially, capturing the microthermal effects is experimentally very challenging. There is considerable interest in use of biomolecular functional materials in areas of molecular detection, energy storage and conversion, adhesive, force dynamics, etc. some of these 2D advanced functional materials (organic and inorganic) have been studied in terms of the EMR response (microwave, THz, and optical frequencies) and behavior of their electromagnetic functions (Zhang et al., 2019). Use of low-dimensional EMR functional materials for biological shielding materials and applications of electromagnetic imaging are described in a review article (Cao et al., 2019). Such an approach using biomaterials will be very useful for considering design of functional materials using biomolecules. External stimuli on mammalian cells respond with oscillatory signaling process (Cheong and Levchenko, 2010). A review on understanding the dynamics of biological and neural oscillator network is published (Bick et al., 2020). Zhang et al (2019) have provided much information about conversion of visible and NIR light into collective plasmonic oscillations in materials such as molybdenum sulfide ($MoS_2$) and black phosphorous. Further attempts to study such equidistant oscillation responses of cells to EMR would be a very important area for future studies.

This paper is organized in 5 sections. Section 2 provides the general biological effects on non-ionizing radiation and incorporates cell responses and impacts for EMR with sub-section 2.1 presenting review of biological effects in the ELF to 300 GHz range that includes mmW frequencies (30GHz-300GHz), sub-section 2.2 provides reviews for biological effects of EMR interaction at the mmW to far infra-red (FIR) band (300GHz-20THz). Subsequently, sub-section 2.3 provides biological effects noted for the FIR to near-visible (NIR) band (20THz-461 THz) including an overview of the low light therapy (LLLT) techniques that are used in the medical industry. Some of their consequences of patient exposure are discussed. Section 3 provides the human safety standards adopted, units used for each frequency domain are explained, and an overview of the fundamental mechanisms (dielectric properties based) governing interaction of THz radiation with biological materials is done. Section 3 also presents some important result summaries of experimental and theoretical works, as well as providing a reference to some of the tools to identify THz thermal/non-thermal effects at all levels of biological organization. This is followed by section 4, which provides an overview encompassing some details and experimental evidence of biological effects studied at ELF to very high frequency regimes, such as in the 3Hz-384 THz range accompanied by an illustration of the frequency effects covering the entire non-ionizing radiation range (Figure 7) with a companion Table 3. The concluding section highlights all the important points mentioned in the preceding sections, with discussions of the limitations of the scattering dominated by the near IR (NIR, 214THz-384 THz), Mid IR (MIR, 100THz-214THz) and the Far-IR (FIR, 300GHz-100 THz) based laboratory studies on biomaterials. Some important theoretical studies and modeling efforts have also been accomplished by specific groups in the NIR window (222THz to 461 THz) for modeling of optical properties of materials at NIR frequencies, which have been discussed. This section also provides some details on new directions and new ideas proposed by researchers that bring more detailed physico-chemical reasoning and understanding of EMR-biological material interactions.

2. **General biological effects of non-ionizing radiation**

A large variety of mmW techniques extant for human safety and communication applications such as the use of airport security cameras, automobile collisions-avoidance radar systems, projected frequency use of 5G and 6G cellular communication framework, crowd control weaponry, and use in medical devices, human cells are expected to be more involved in thermal effects and undergo continuous resonant interaction of EMR with the membrane proteins. A systematic study on the relationship between mmW on the function of key proteins involved in neuronal excitability and neural signaling was reported (Shapiro et al. 2013). Authors have found significant acceleration of kinetics, and an increase in action-potential firing rate of oocytes due to the combined effects on sodium and potassium channels thermally mediated effects on excitable cells; thermal effects are pronounced, and authors report that power ranging between 4 mW- 128 mW irradiation resulted in acceleration of kinetics and/or increase in activity. Continuous wave (CW) mmW exposure of 100 s produced maximal temperature elevations in range 1.1°C ± 0.2°C to 8.9°C ± 1.5°C. Using a 60 GHz probe irradiation technique, they expressed voltage-gated sodium and potassium channels, as well as sodium-potassium pump in Xenopus laevis oocytes. They report considerable activity observed in 3 key proteins. They also suggest that further studies may be required to deconstruct the thermal aspects of mmW response in more complex systems, such as innervated cell cultures, and intact organisms. There is a conjecture that the specific biomolecule responsible for neuronal excitability may have different thermal coefficient ($Q_{10}$) for different species. The authors have also strongly suggested that with the use of a broader frequency range of the irradiating EMR beam, various other thermal effects may be possible to be revealed.

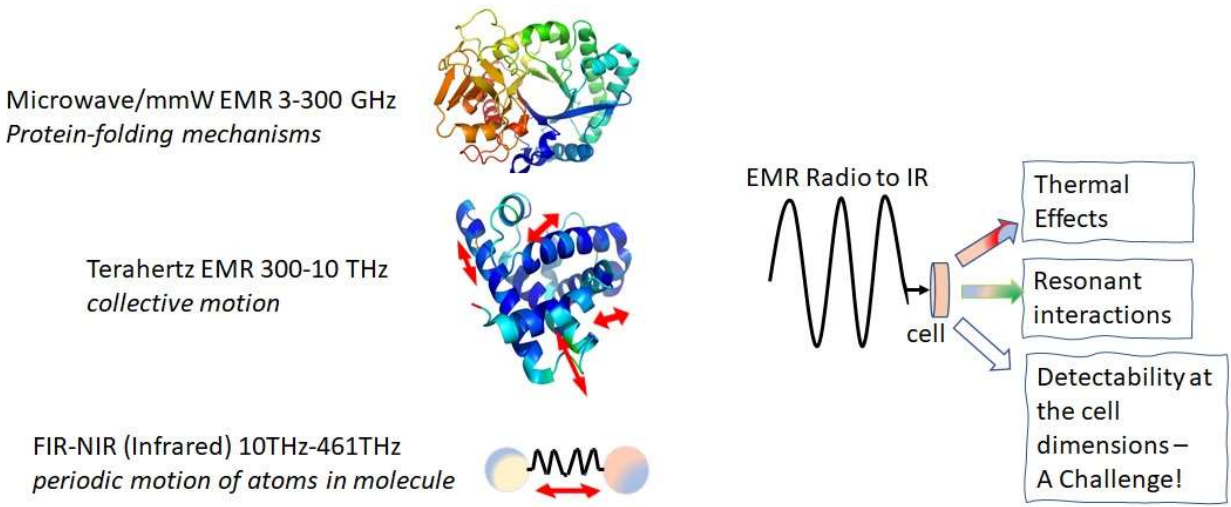

**Figure 2.** Cellular level phenomena shown in this schematic for microwave (protein-folding), THz (collective motion) and IR (atomic vibrations) with three main challenges: thermal effects, resonant interaction, and detectability inside the cell at the cellular dimension. This figure also has pertinence to Table I given in section 2.2.

Figure 2 shows the 3 principal mechanisms through which cells interact with EMR in the range ELF-NIR domain. It is pointed out that thermal mechanism underlies the effects of mmW on both the firing rate and the ionic channels of *Lymnae* neurons, and they also show that many ionic channels are reversibly modified because of thermal effect on mmW and microwave radiation (Alekseev and Ziskin 1999). Kucera and Cifra (2016) have investigated the nanoscale interaction of EMR with bio-macromolecules. They provide details about the micro- and nanoscale characterization of both passive and active RF properties of bio-macromolecules and cell, experimentally determined viscous damping of structural vibrations, and performed detailed analysis of energetic circumstances of EMR between oscillating polar species. It is pointed out that the detectability of EMR radiation field in the scales conforming to the size of cells is still not achieved. Techniques for such detection levels must improve for future work.
12

## 2.1 Biological effects in extremely low frequency (ELF) to mmW band (3Hz-300 GHz)

An in-depth review of the static- and time-varying non-ionizing radiation fields in the extremely low frequency (ELF) range 3 Hz to 3kHz and its interaction with the biological systems is well presented in the International Agency for Research on Cancer (IARC) monograph on the evaluation of carcinogenic risks to humans (IARC 2013). Three types of epidemiological studies are reported in this monograph involving cohort-, case-control and correlation paradigms. Cohort and case-control studies provide relative risk factors (ratio of incidence to actual mortality) of cancer in humans establishing a direct relationship between exposure and cancer occurrence in individuals. For correlation studies, units of investigations are entire population in a geographical area where the exposure probability is quite high. In the ELF regime, the static electric, and magnetic fields (man-made) are of the order of 10-100 V/m and 0.1-1.0µT, respectively. It seems that the man-made magnetic field intensities are $1/50^{th}$ of the natural geomagnetic field strengths. However, there could be transients in the ELF ranges and might contain a lot more frequency than 3 GHz. At ELF frequencies the electric and magnetic fields are generally weakly coupled and at these frequencies the size of humans and animals are very small compared to the wavelengths. The static electric fields being always oriented perpendicular to human body induces surface charge density. Static magnetic fields interact with tissues by direct electrodynamic process with ionic currents, such as nerve signal or blood flow, and induces electric fields across the blood vessels. The second type of interaction is magneto-mechanical and results in orientation of biological structures under the strong magnetic fields. The third way applied EMR magnetic fields impact biomolecules is through the Zeeman effect which is responsible to change the molecular energy levels often leading to changes in concentration of free radicals that may become highly reactive. However, at ELF frequencies the photon energy is so small (~ $10^{-12}$ of the weakest chemical bond energy) direct interaction such as chemical bond breakage, and DNA damage is very low. Based on a cohort study for childhood cancer and exposure to ELF magnetic fields (calculated historical magnetic fields), it is mentioned in the monograph that the standardized incidence ratio (SIR, ratio of number of observed to expected occurrences) for magnetic fields < 0.01 µT SIR= 1.0, for 0.01-0.19µT, SIR = 0.94 and for fields ≥ 2.0 T, SIR=1.2. This is based on a sample size of 68,300 boys and 66,500 girls under the age of 20 years. Case-control studies (odds ratio) of childhood leukemia and childhood tumor of the central nervous system and exposure to ELF magnetic fields is exhaustively presented in Table 19 and 20 respectively, in the Monograph. Table 21 of the Monograph gives the case-control study-based risk estimates (OR, odds ratio) of childhood leukemia and exposure to ELF electric fields that provides data based on a 1999 personal monitoring based on 274 subjects with 331 controls aged 0-14 years as for E < 1.2 V/m, OR=1.0, for E between 12.2-less than 17.2 V/m, OR=0.79, for E between 17.2- less than 24.6 V/m, OR=0.76, and for E between 24.6-64.7 V/m, OR=0.82.

Before the second world war much of tissue electrical properties were documented such as capacitance, permittivity of many types of cell suspensions and tissues were characterized for EMR frequencies up to 100 MHz (Topfer 2015). Measurement based inferences were drawn for the cell impedances (a frequency dependent complex resistance parameter). Just after world war II when intense advancements were made in source, detector and measurement technology, systematic measurements of dielectric properties of different tissue types were achieved up to 10 GHz (Topfer 2015).

Non-thermal microwave effects on biomolecules are not discernible in low electric fields less than $10^7$ volts/m. Barak et al. (2009) studied the fluorescence emission of the enhanced green fluorescent protein (EGFP) in aqueous solution under CW microwave irradiation in a transverse electric $TE_{011}$ (high Q) cavity at 9.5 GHz. They note that the dominant microwave irradiation effect on EGFP is the temperature rise associated with the microwave heating. In the fluorescence spectra, they found that the microwave-induced effect is purely heat producing in the 500-540 nm band, while the 540-560 nm effect is non-thermal and authors attribute this to microwave-induced reduction of fluorescence anisotropy (lifetime reduction) and terms its "vibronic" transition.

Yang et al. (2015) have studied use of structure resonance energy transfer (SRET) for H3N2 virus inactivation using a threshold electric field of EMR in the range 6-12 GHz. This technique uses confined acoustic vibrations (CAVs)They have reported quantitative agreement of experimental results on threshold electric field estimates with those of a well-conceived model that uses dipolar mode of a homogeneous sphere(considering virions of influenza virus as spherical balls) and then using microwave EMR to excite dipolar resonance of the virus. They observed a strong resonance that led to virus inactivation at around 8.4 GHz due to the proposed SRET. In this experiment authors also reported that they have validated the virus inactivation ratio without causing any RNA genome using the real time reverse transcription polymerase chain reaction (RT-PCR) technique and found the virus inactivation ratio increases with EMR power density.

In a recent experiment (Roy et al., 2020) an X-band (10.525 GHz) low photon energy (0.04 meV) irradiation apparatus was used to expose 4 sets of prostate cancer (PCa) cells. The experimental arrangement (schematic) and the EMR transmission (%) are shown in Figure 3 below. The goals were to note the effect of low radiofrequency EMR on prostate cancer (PCa) cell morphology and oncogenic cell signaling. About one million transformed-immortalized human prostate cell line RWPE-2 cells and prostates cancer (PCa) cell lines PC3, LNCaP (Caucasian origin), or MDAPCa-2B (African American origin) were exposed for 120 minutes with low radiofrequency EMR and cells were further grown for 16h at 37°C. Results suggest that no major morphological changes were observed however, exposure increased expression of phosphorylation of serine 2448 of mammalian target of rapamycin (pS2448-mTOR) in RWPE2 and MDA-PCa2B cells and phosphorylation of serine 473 of AKT kinase (pS473-AKT) in LNCaP and PC3 cells.

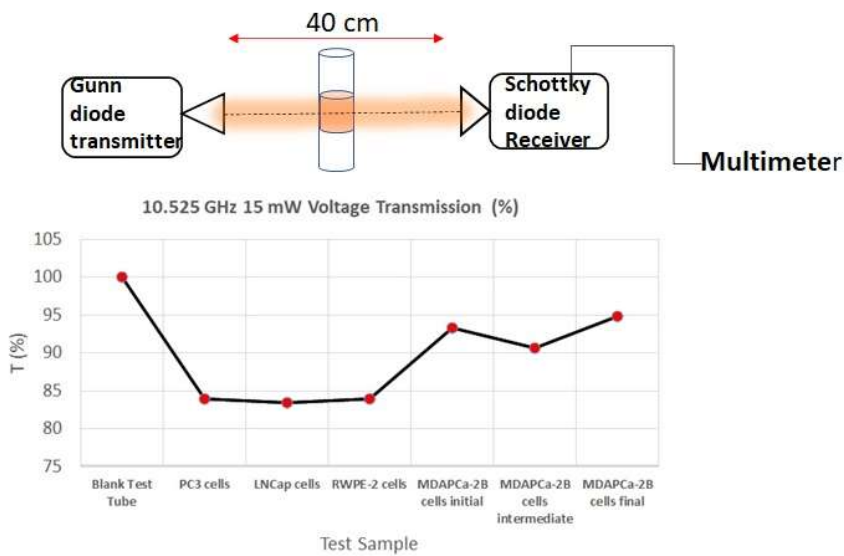

**Figure 3.** Schematic of the X-band RF irradiation setup to study normal, African American and Caucasian prostate cancer cells with PASCO 10.525 GHz (2.85 cm) transmitter with 15 mW beam, photon energy ~ 0.04 meV and sustained CW exposure time 120 minutes

Since phosphorylation of mTOR and AKT enhances the aggressiveness in PCa cells, we further analyzed the expression of oncogenic molecule tumor necrosis factor alpha induced protein 8 (TNFAIP8) and an apoptotic marker cleaved PARP (cPARP) expression in PCa cells after exposure of 10.525 GHz EMR. Our data suggest that TNFAIP8 expression increased in all PCa cells after exposure, whereas the expression of cPARP is downregulated in RWPE2 and LNCaP cells suggesting this might induce endogenous oncogenic signaling in PCa cells by modulation of AKT/mTOR and TNFAIP8 expression. Figure 4(a) shows the data obtained after normal prostate cancer RWPE2 cells and MDAPCa-2B, PC3 and LNCaP prostate cancer cells (one million) were exposed with the 10.5 GHz EMR for 2h and cells were further grown for 16 h at 37°C. Cell lysates were prepared as described previously (1) and fifty micrograms cell lysates were immunoblotted with indicated antibodies.

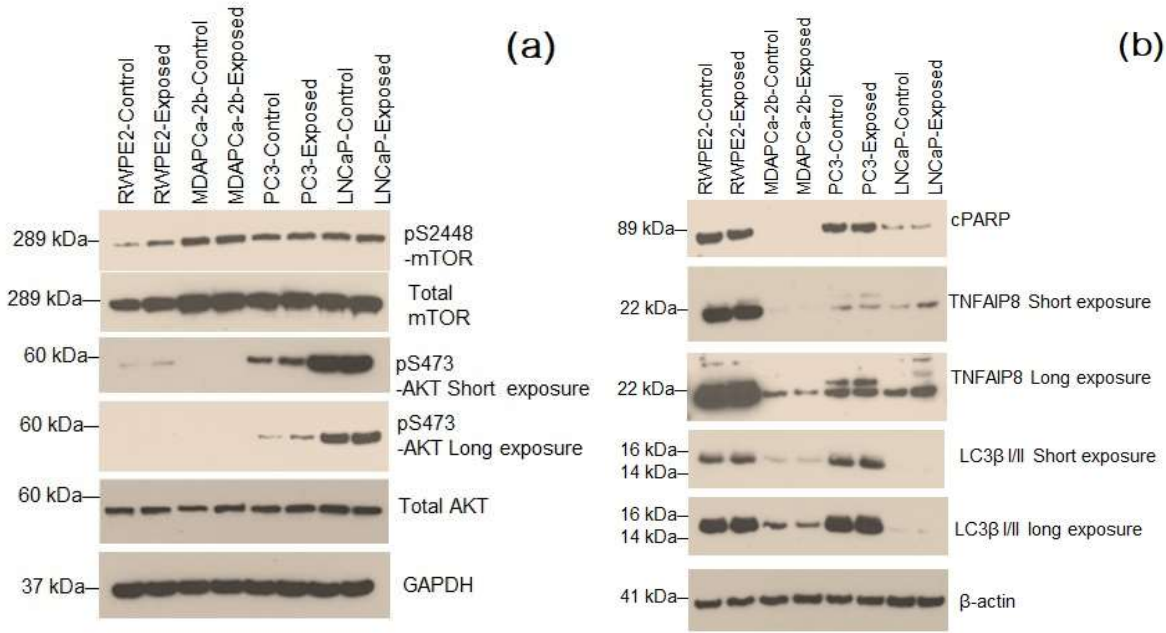

**Figure 4.** (a) Exposure of 10.5 GHz EMR activates AKT/mTOR signaling by increasing phosphorylation of AKTS473 and mTOrS2448 in normal and prostate cancer cells indicates that, 10.5 GHz EMR may promote cell proliferation not only in normal prostate cells but also in prostate cancer cells. Protein masses expressed in unit kiloDalton (kDa), (b) Exposure of 10.5 GHz EMR decreased cPARP levels compared with control cells suggesting that decreased in cell apoptosis. However, exposure of the X-band (10.5 GHz) EMR increased the expression oncogenic molecule TNFAIP8 which promotes prostate cancer cell growth. No major change in autophagy marker LC3β I/II expression was observed after EMR exposure to the normal and prostate cancer cells.

Figure 4(b) above gives data (Roy et al., 2020) on normal prostate cancer RWPE2 cells and MDAPCa-2B, PC3 and LNCaP prostate cancer cells (one million) when exposed with 10.5 GHz EMR for 2h and cells were further grown for 16 h at 37°C. 50 microgram of cell lysates were immunoblotted with indicated antibodies.

Ambient humidity plays an important role in the manner heat is deposited on ocular cells of humans (Kojima et al. 2019). Higher ambient humidity results in stronger heat accumulation and transfer leading to ocular temperature elevation. These experimental evidence were established by systematically collecting ocular tissue temperature rise once when the humidity was 70% and another time when the humidity was 30% and using 40 GHz EMR with power density 200 mW/cm$^2$ for 5- or 30-minute exposure time. All experiments were done in room temperature, and dynamic changes of temperature and its distribution in anterior chamber, saline containing 0.2% microencapsulated thermochromic liquid crystal (MTLC) was injected into the eye chamber.

In the modern era, a lot of microwave-millimeter wave devices are being used for diagnostic applications in medical facilities. An excellent review work has been done that presents the evolution of human body-centric wireless applications, and implications on the cellular level biology (Zhadobov et al. 2011) for frequencies in the millimeter wave (mmW) spectrum (30-100 GHz). EMR, thermal, and biological aspects are reviewed. It is documented that 26%-41% of the EMR power is reflected from the air-skin interface for normal incidence, more than 90% of transmitted power is absorbed by the skin, hence, a single layer skin model is enough for EMR dosimetry applications. The clothing in direct contact with skin enhances EMR transmission whereas the air gap (~0-2mm) between clothes and skin impedes the transmitted power. An exhaustive review is done for the results of *in vitro* and *in vivo* biological exposure studies and categorized for the low- (< 5 mW/cm$^2$) and high (> 1W/cm$^2$) power density exposures. Authors have identified 3 significant biological effects of mmW interaction with cells, viz. effect of cellular proliferation, effect on genetic expression, and effect on bio-membranes. Authors have identified that 52-78 GHz EMR exposure reduces proliferation of human melanoma cells. Cell proliferation is not affected by low power mmW, anti-cancer properties of mmW may be indirect, and the exposure to EMR can reduce tumor metastasis through activation of "killer" cells. On the effect of genetic expression, authors point out that due to non-ionizing nature of the mmW they are not genotoxic (production of chemicals that damage the cellular DNA) but possibly exhibit proteotoxicity (adverse effects of damaged proteins). The authors also mention that they bypassed the mmW thermal effects and studied the change of protein expression such as HSP70 (a stress induced protein) but could not substantiate the change. Authors analyzed reticulum stress data based on 0.14 mW/cm$^2$ 59-61.2 GHz radiation and they did not observe any effect on endoplasmic reticulum homeostasis, hence, conclude that mmW does not trigger an acute stress. Structural changes in cellular membranes were observed due to exposure at 42.2 GHz at a power density 35.5 mW/cm,$^2$ however, no cellular damage was observed. Authors point out that that such structural modifications may play a role on cellular signaling/interactions. With 130 GHz pulse modulated EMR, and with 53 GHz radiation, the cells can experience physical changes that can modify permeability of phospholipid vesicles.

Pattanaik (2012) has presented a general review work on the biological effects of RF or microwave EMR on humans and provide a critical discussion of the cause and effect of thermal and non-thermal mechanism, resulting from human exposure. This author has mentioned that thermal mechanisms directly deposit RF power in the biological system. Pulsed microwave EMR generates acoustic transients in the tissue due to transient heating of the water and the simultaneous thermal expansion. The paper mentions that the human head exposed to RF energy with carrier frequency ~ 1 GHz and peak intensities ~ 10KW/m$^2$ can increase the tissue temperature by a few micro-degrees Celsius but at a very high rate. This will lead to acoustic transients in the head more than 10 dB acoustic equivalent power level. The non-thermal mechanisms are due to electric fields that exert forces on charges and magnetic fields exerting torque on magnetic dipoles in biological tissues. This results in shocks and membrane excitation. Electric fields can create pores in cell membranes by producing electrical breakdown. Membrane potential ~1 Volt requires tissue field strengths exceeding $10^5$ volts/m. Electric fields also exert forces on ions and torques on permanent dipoles (dielectrophoretic forces).

Simulation results using appropriate models representing 4 different body parts, exposed to 40-100 GHz energy have been documented (Wu et al. 2015). They have found, at 60 GHz the power reflection coefficient may vary between 34% and 42% at the air/skin interface for normal incidence due to the variation in dielectric parameters. They recommend that for thermal modeling, a multi-layer skin model is more appropriate.

A 2016 study related to 24-hour exposure of human corneal epithelial (HCE-T) and human lens epithelial (SAR01/04) cells to 60 GHz (Koyama et al. 2016) reveals that exposures of HCE-T and SRA01/04 cells has no significant effects on the micronucleus (MN) formation frequency. They conclude that such exposures do not have adverse effects on the genotoxicity of Hsp expression of cultured HCE-T and SRA01/04 cells. However, they have suggested to repeat the study for other EMR frequencies to confirm this evidence.

Rat glial cells were systematically exposed to mmW EMR (power density 3.2 mW/cm$^2$) in the frequency range 120 to 180 GHz. Authors found that after 1 minute of exposure, a relative number of apoptotic cells increased 1.5 times and after 3 minutes, the numbers doubled (Borovkova et al. 2017). They outlined, that since many of the intracellular structures of glial cells are significantly polarized, and the size of glial cells is less than the wavelength of the mmW, a direct resonance absorption is somehow possible resulting in cell death. They also attribute mmW absorption by water (culture of glial cells in aqueous medium) leading to the change of state of the membrane protein molecules.

A long term (28 day) exposure of 1.8 GHz EMR on mice did not show significant effect of depression-like behavior (Zhang et al. 2017), spatial learning and memory ability or the histology of the brain. But they found that the level of anxiety enhanced, and possibly also increase of the amino acid neurotransmitter such as GABA. Anxiety like behavior in this study were evaluated by different tests such as the sucrose preference test (SPT), tail suspension test (TST) and forced swim test (FST). Spatial learning and memory ability tests were evaluated using Morris water maize (MWM) procedure. Amino acid neurotransmitter levels were determined by the liquid chromatography-mass spectrometry (LC-MS) and the histology of the brain was examined by the hematoxylin-eosin (HE) straining.

Currently, restrictions on human exposure to EMR waves at frequencies higher than 10 GHz are generally defined by the incident power density, to prevent excessive thermal effects in superficial tissue (Laasko et al. 2017). They concluded using finite-difference time-domain (FDTD) modeling of the bioheat transfer equation, and investigating the effects of frequency, polarization, exposure duration and depth below the skin surface to the temperature rise for the human face when pulses are shorter than 10s at EMR frequencies between 6-100 GHz. These authors point out that the current pulsed radiation exposure limit standards and guideline are not properly evaluated, and as such suggested a revised method of estimating the temperature rise of the human face for pulsed radiation (shorter than 10 s) by including the actual energy absorption per unit exposed area (radiant exposure).

Based on the reviewed papers for work reported in the ELF to 300 GHz band, it is noted that maximum number of experimental works was accomplished in the 30-100 GHz band. Figure 5 is a bar plot showing the number of publications made in each sub-band.

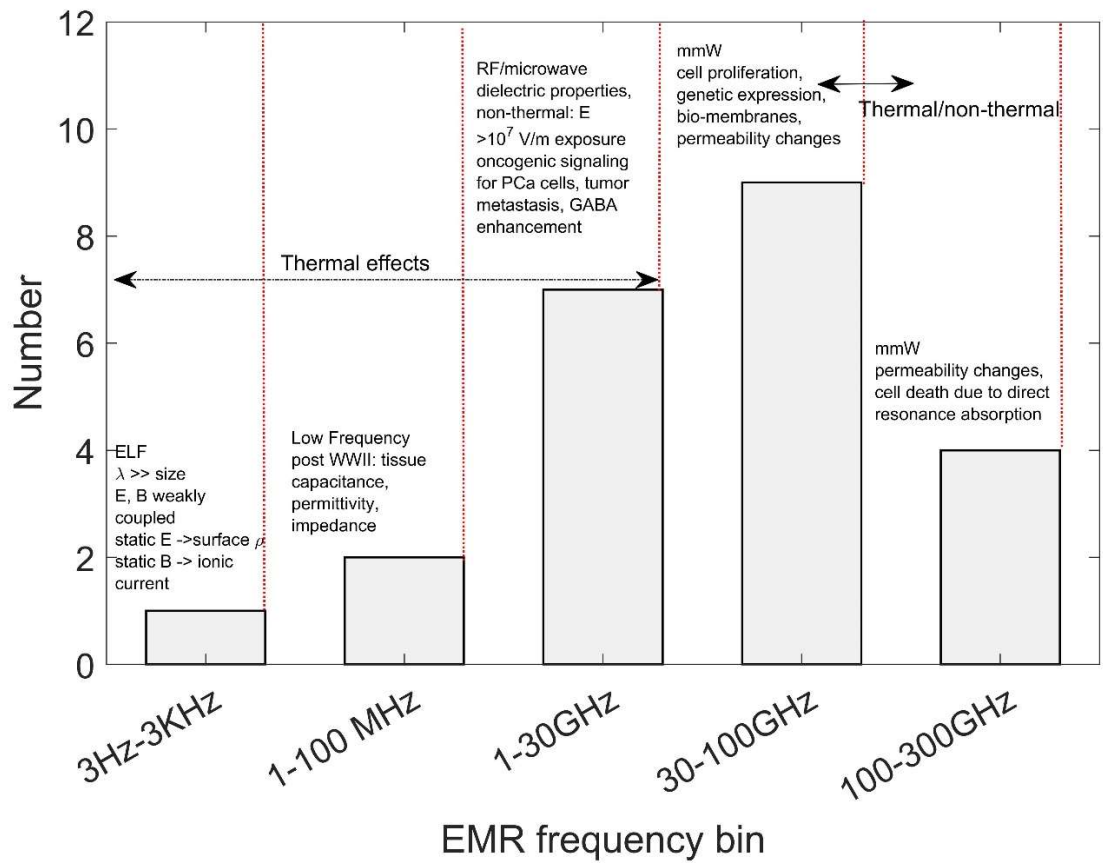

**Figure 5.** Bar plot showing the number of research work reported in the ELF-300 GHz with dominant physical mechanisms, and significant bio-effects. Typical properties in each sub-band and important bio-effects are mentioned for each investigative sub-band.

### 2.2 Biological effects and investigations for mmW to THz/FIR band (300 GHz-20 THz)

For studying exposure of cells and tissues and finding the effects, the "terahertz" (THz) domain used in literature extends between 300GHz to 20THz. The complete spectral information of registering biological effects is an open field of research since not much exposure/biomodulation and impact related data has been acquired for reasons of measurement difficulties such as cost of equipment, not so easily available THz EMR sources, detector, and wideband digitizer systems.

Nonetheless, interaction between THz EMR and biological tissue work at fixed frequencies has been going on for some time at various laboratories throughout the world. Many interesting review works have been published. A review (Zhao et al., 2014) points out the following uniqueness of THz EMR:

- THz EMR in the range 300GHz-3 THz has seven strong merits viz.
- THz pulse widths (sub-ps scale) allow for facile analysis of various materials, A few cycles of oscillation of the THz source (pulsed-mode) provides the frequency range in each pulse in the range GHz-10THz,
- THz radiation is highly coherent in space and time hence, this property enables very accurate estimates of dielectric properties of material/biomaterials.
- THz photon has so low energy level that prevents ionization hence, highly applicable for biopsy studies.
- THz EMR has very strong penetration power for non-polar materials thereby very good detection ability.
- Polar molecules strongly absorb THz EMR and this feature can be used for finished product quality control non-destructively.

In 2014, authors felt that studies related to biological interactions of THz EMR were not accomplished as much, and suggested cross-integration involving physics, biology, materials science and medicine and sensor electronics development.

An Institution of electrical and electronics engineering (IEEE) invited paper on selected topics in quantum electronics (Markelz 2008) is a very important review work on THz dielectric response measurements on large biomolecules using THz dielectric spectroscopy techniques. This is a relatively a new field and requires attention. In this context, the author has given a very detailed explanation of relative contribution of the dielectric parameters of the large biomolecules between the vibrational and relaxational modes. Relaxational modes are explained as the time-consuming process of adjusting the relative redistribution of charges to the field lines exerted by the probe EMR. The author has mentioned that when sample uniformity is maintained and multiple reflection effects are included properly in the experimentally acquired data analysis, the overall dielectric response of large biomolecules in the THz range of frequencies is broad and featureless. The author also concluded that the response is very sensitive to the level of hydration, temperature, binding, and conformational changes. In biological molecules, the amino acid side-chains exhibit interaction of the applied probe field through weak hydrogen bond and respond naturally by aligning the local dipole. Authors also point out that the overall dielectric response of large biomolecules can be attributed to surface sidechain relaxational loss, which they investigated concerning the environment, structure, and function.

Long term THz exposure data related to skin absorption (0.3-1.5 THz) are highly variable (Vilagosh et al. 2019) and such data are very scanty. These authors have attempted to standardize the measurement of skin absorption at diverse temperatures and devised a method utilizing the complex permittivity $\varepsilon^* = \varepsilon' + i\varepsilon''$, $\varepsilon'$ being the real part and $\varepsilon''$ being the imaginary permittivity that affects the absorption magnitude and all parameters as function of THz frequency. They report complexity involved in ascertaining the absorption from the "bound water" to biological molecules. The presence of bound water in biosystems influence the complex permittivity in a manner equivalent to reducing the temperature. The lowering of $\varepsilon''$ but not the real part, $\varepsilon'$ with increasing concentration of polar compounds that produce greater concentration of bound water has been demonstrated by many investigators. Bound water is also released in cell death by the process of protein degradation and has been verified and it is mentioned in a cross reference (Huff-Lonegran and M. Lonegran, 2005) that much of water in muscles are entrapped in cell structure that includes intra- and extra myofibrillar spaces. Changes in intracellular architecture enables the cell retain water. Skin that has been degraded is expected to have these bound waters released, thereby enhancing the imaginary ε" for a given temperature – this complicates analyses. However, authors have also suggested that human skin is a complex, multilayer variable, and cannot be just simply understood by modeling the water content, but skin components such as melanin and collagen structure in the dermis must also be accounted for that influence THz absorption process.

Water in skin absorbs THz EMR very strongly hence, penetration depth is only a few hundred microns of the skin tissue resulting in a great opportunity for the *in vivo* skin studies and routine examination procedures. Lindley-Hatcher et al. (2019) have pointed out that variation of pressure with which the skin region is imaged (contact pressure on the imaging window) has a very sensitive response on the THz image signal because the shift in apparent density due to the external pressure distorts the water distribution (strong absorber) and hence, the repeatability of the procedure is hindered. Authors have proposed a robust and rigorous protocol to obtain a repeatable result using a pressure sensor device to control the contact pressure while processing the skin reflected THz signal by applying a double Gaussian filter to remove high- and low frequency noise.

THz spectra of biotin (an important water soluble polar and optically active biomolecule that has low absorption lines) were simulated using first principle, and density functional theory (DFT) using harmonic approximation, and is sensitive to entire spectrum related to the atomic vibrations (Bykhovski and Woolard 2013). THz technology is being widely accepted for real time data acquisition at tissue level while a patient is under surgical procedures. Human blood was investigated for detection of real time tissue pathology during surgical procedures. THz signal from tissues is generally convolved with those resulting from EMR interaction with the biomaterial, fluids such as blood, hence THz time-domain spectroscopy (TDS) technique was applied to quantify the frequency-dependent absorption coefficients, refractive indices, and Debye relaxation periods of whole blood, red blood cells, plasma, and thrombus separately (Reid et al. 2013). The authors have reported measured absorption and refractive indices of whole human blood, blood-plasma, and blood cells for the frequency 0.2 to 2 THz. They also report that the measurement taken at THz frequencies are sensitive to the bonding mechanisms of the whole blood. For breast cancer imaging (in breast conservation surgery) a 1.89 THz system was built to locate cancer. The equipment was operated in reflection mode. Images were correlated with the optical histological micrographs and a mean discrimination of 73% was noted (St. Peter et al. 2013). In this paper, authors have provided the refractive index (n) and power reflectivity by calibration fluid and the tissue type using the reflection coefficient (Γ), $|\Gamma|^2 = \left(\frac{n-1}{n+1}\right)^2$ and they also report that the THz power reflection is predicted 20% higher for carcinoma than for regular fibrous tissues. Hydrofluoric acid burns in compact bone (osseous) tissues were successfully obtained using the THz-TDS technique. Various image reconstruction techniques were tried to improve the contrast in the features of interest (Baughmann et al. 2013). Authors in this paper reports that formalin used for histological fixing causes contrast between damaged and undamaged tissue. However, the THz was not generated solely from a source in this experiment, rather, electro-optic sampling was employed on a 120 fs, 790 nm ultrafast laser beam with a repetition rate of 86 MHz and signal was acquired in the time domain, and then fast Fourier transformed to resolve the time domain signal to a signal form having a frequency resolution of 25 GHz.

Wavelengths of EMR (RF current in tissues) are much longer than depth of human structures hence, quasi-static electric field treatment is applicable (Paruch, 2020). The RF tissue heating process can be modeled using Boltzmann transport equation for thermal transport (in electric field) using current density ($\rho$, in Coulombs m$^{-3}$) and energy density (u, in Joules m$^{-3}$) equation of continuity

$$\frac{\partial \rho}{\partial t} + \nabla \cdot \boldsymbol{j} = 0 \qquad (5)$$

$$\frac{\partial u}{\partial t} + \nabla \cdot \boldsymbol{j_u} = \boldsymbol{j} \cdot \boldsymbol{E} \qquad (6)$$

In 3D Cartesian coordinate, the del operator $\nabla = \left(\frac{\partial}{\partial x} + \frac{\partial}{\partial y} + \frac{\partial}{\partial z}\right)$, **E** is the electric field intensity (Volts m$^{-1}$), **j** is the current density in sample (**j** = $\rho$**v**, **v** being the velocity of charges at the point) in Amperes m$^{-2}$, **j$_u$** is the energy current density based on the semiclassical dynamics in Boltzmann transport distribution function

$$j_u = 2 \int_\Omega \frac{d^3 k}{(2\pi)^3} u \boldsymbol{v} \delta f \qquad (7)$$

$k$ is wave-vector (spatial angular frequency of the interacting RF wave in tissue signifying how many oscillations made in the interacting shell volume $d^3k$), f being the distribution function, and $\Omega$ representing the first Brillouin zone. To draw the connection, first we relate the energy density time derivative to the specific heat at constant volume ($c_v$) in local thermodynamic equilibrium condition (LTE)

$$\frac{\partial u}{\partial t} = \frac{\partial u}{\partial N_e} \frac{\partial N_e}{\partial t} + \frac{\partial u}{\partial T} \frac{\partial T}{\partial t} \qquad (8)$$

We can write for u=u(T,v) and constant volume process for the LTE and apply the chain rule,

$du = \left(\frac{\partial u}{\partial T}\right)_v + \left(\frac{\partial u}{\partial v}\right)_T$, second term yielding zero due to constant volume process. Applying the first law of thermodynamics and using $du = \delta Q - pdV$, Q being the amount of heat transferred to the tissue, we can equate, $\left(\frac{\partial u}{\partial T}\right)_v = \left(\frac{\partial q}{\partial T}\right)_v$, where $c_v = \left(\frac{\partial u}{\partial T}\right)_v$, The RHS first term in Eq. (8) can be simplified using $N_e = -\rho/q_e$ ($q_e$ is the electronic charge), and using $du \approx \sum_i \mu_i \, dN_{ei}$ under LTE and constant, µ being the chemical potential of the particle in the system representing how the Gibbs function will change with composition of the tissue, $N_e$ being the number density ($-\rho/q_e$) of electrons (m$^{-3}$), $q_e$ being the electronic charge in Coulombs, and $c_v$ is specific heat of the biological sample at constant volume, Eq. 8 can be rewritten

$$\frac{\partial u}{\partial t} = -\left(\frac{\mu}{q_e}\right)\frac{\partial \rho}{\partial t} + c_v \frac{\partial T}{\partial t} \qquad (9)$$

Using Eqns. 5, 6, and 9

$$c_v \frac{\partial T}{\partial t} = \frac{\partial u}{\partial t} + \left(\frac{\mu}{q_e}\right)\frac{\partial \rho}{\partial t} = j \cdot E - \nabla \cdot j_u - \left(\frac{\mu}{q_e}\right)\frac{\partial \rho}{\partial t} = j \cdot E - \nabla \cdot j_Q \qquad (10)$$

Where $j_Q = j_u + \frac{\mu}{q_e} j$ is the heat current-density that can be related also to the Peltier coefficient $\Pi$ and the gradient of Temperature, $j_Q = \Pi j - K\nabla T$, K being the thermal conductivity. Using the identities $j \cdot E = \sigma |E|^2$ and $\nabla \cdot (\nabla \emptyset) = \nabla^2 \emptyset$, we can write the RHS of Eq. 10 for steady state case (Poisson's form) and zero electric current

$$\nabla^2 T(x, y, z) = -\left(\frac{\sigma}{k_T}\right)|E|^2 \qquad (11)$$

Using Eq. (11) a mathematical model (Neelakanta and Sharma 2013) simulated the electrical conductivity ($\sigma$) to dielectric constant ($\varepsilon_r$) of wet tissue in the 0-10 MHz range, microwave endometrial ablation (MEA) for using EMR between 0.9-2.45 GHz, skin depth of wet tissues in the 0-250 GHz range, and THz endometrial ablation (TEA) related surface temperature rise due to EMR power in the frequency range 30-350 GHz compared with the microwave temperature rise (per second) at 2.54 GHz. This computational study addresses dysfunctional uterine bleeding by shallow ablation technique. Authors have pointed out that designing the effective antenna with exactly matched standing wave ratio is physically challenging. Using Eq. 11, authors have derived expression for the near-surface temperature of the biological medium as function of its heat capacity, electrical conductivity (or magnetic intensity) and the EMR electric field intensity (E). This paper is a good feasibility study based on theoretical approaches for use of microwave and THz EMR for endometrial ablation techniques, possibly good for future clinical strategy.

A review on experimentally concluded THz radiation effects and biological applications was published (Orlando and Gallerano 2009). Authors point out that the spectral region 100 GHz to 10 THz represents an opportunity for research in physics, chemistry, biology, and medicine. They came up with the idea of incorporating "evolved molecules" in biological network systems that involve highly specific interactions of molecular complexes and using a term to collectively call them as part of a nanomachine that would require systematic studies for a large range of time scales from sub-picoseconds to an ultraslow conformational relaxation period. They point out that THz science enables analysis of fast protein vibrations, that would help us better understand the physical basis of enzyme catalysis, thereby necessitating studies on proton and electron transfer processes. They also point out that bulk water is still difficult to understand in the THz domain especially the biomolecular coupling with water. Further work in THz spectroscopy might prove to be very beneficial to answer these basic questions. Many experimental approaches are discussed, and suggestions are made by the author while discussing the outcomes of the European Union (EU) THz -BRIDGE) experimental reports (2001-2004) and ongoing THz-BEAM studies. EMR exposure of proteins are a prime concern for quantifying human exposure. The exact fate of THz absorption by cells are still unknown, however, the damages incurred are not permanent like those induced by ionizing radiation such as X-ray. Authors have discussed the manifold effects of THz radiation, such as on cell membrane, the electric fields are normally $10^5$V/cm and exhibits very strong polarization; non-thermal effects for high frequency EMR (at low intensities) is an area of controversy. Cell permeability under THz exposure for liposomes were studied using pulsed 130 GHz EMR at 5-7 Hz with time-averaged power density 17 mW/cm.$^2$ Liposomes are about 4 nm thick resulting peak fields are 2 orders less than those of lipid bilayers. With the electric field window between 2.6-2.7 KV/m a possible rectification of the THz pulse by liposomes has been hypothesized. This could result in change of functionality of lipid membrane as such. Authors also point out that genotoxic effects are induced by THz EMR and DNA damage of somatic cells lead to cancer. Abnormal chromosome number (aneuploidy) is an indicator of genomic instability for the cell to respond the EMR. Asynchronous replication of centromeres showed increased levels in lymphocytes when exposed to 100 GHz EMR with power density ~ 31 μW/cm.$^2$ Authors suggest that a possible mechanism may be the ability of EMR to excite low frequency collective vibration modes for biomolecules, these are capable to produce resonance excitation as well, that might lead to mutagenesis.

THz pulse propagation in biological systems in the range of 100 GHz-3 THz are used for medical imaging. Pickwell et al. (2004) published a theoretical paper dealing with their development of a finite-difference-time-domain (FDTD) simulation model for studying propagation of THz pulse through human skin especially highlighting the effect of water embedded in skin structure. They published the comparison table of experimentally determined double Debye parameters for water with their FDTD (applying an *in vivo* model) simulation derived parameters.

THz-TDS is a highly coherent detection technique for understanding the structure and function of biomolecules (Yu et al., 2019). Authors presented an effective THz absorption spectrum in the 500 GHz-10 THz band of 5 typical nucleobases of DNA/RNA. They used a super broadband THz detection technique called the air-biased coherent detection (THz-ABCD) technique. ABCD technique relies on use of 2 color laser induced air plasma and a broadband detector system resulting in more THz power with a broader bandwidth of the source and detector. Authors identified and concluded stating THz-ABCD system has a better response on biological system than the THz-TDS systems in terms of identifying molecular vibrations and twists.

An experimental report was generated on studies related to use of THz spectroscopy technique (100 GHz-10 THz) to explore and diagnose the genic mutation of DNA molecules (Tang et al., 2018). They detected DNA oligonucleotides by combining information from THz spectroscopy and microstructures. Based on this study they suggested that a microfluidic chip as a carrier device helps reduce the strong water absorption of the THz EMR. Oligonucleotides can be differentiated according to respective THz absorption spectra, and also, they exhibit different redshift in resonance frequency because they exhibit different dielectric properties, respectively. Difference in number of hydrogen bonds formed by oligonucleotide molecules and surrounding water molecules can be used to explain THz absorption coefficients.

Additional literature (Crowe et al. 2003) suggests that direct Frequency-Domain THz Spectroscopy (FDTS) may be very useful for investigating the properties of biomaterials while studying the structure and dynamics of biomolecules such as RNA, DNA, and protein. They have shown that bacterial spores have distinct THz spectra, which may be useful for identification of bio-agent. They explained data obtained using all solid-state THz source and detector system operating and measured DNA material spectra in the 0.18-0.8THz.

Brown et al. (2010) published data about narrow THz spectral response (attenuation signatures) of RNA in nanofluidic channels with small interfering RNA samples (25 bp/mol) suspended in TPE (Tris-Phosphate-EDTA) buffer solution. Authors preferred using a time domain spectroscopy (TDS) method employing THz mixing technique which provides a very good signal to noise ratio and regarded as a highly coherent spectroscopic technique. They used a sweep oscillator 0.1THz-2THz with a linewidth ~ 100 MHz and came up with the idea of replacing the conventional THz thermal detectors (such as Golay cells) with the coherent photo mixing transceiver because the thermal detector response is highly limited by the noise equivalent power at room temperatures. After confining the RNA suspension in TPE solution to an array of 600-nm x 500 nm silica nanochannel fabricated on a very high resistivity silicon substrate and performing successive transmission measurements they find consistent signature width (between the -3dB points) ~10GHz. They attribute the cause of such widths due to linearization, surface binding that could enhance the oscillator strength of vibrational modes.

One of the limitations of THz band for nanoscale objects such as cells, molecules etc. is that the wavelengths associated with the radiation in this band is much longer compared to the object size. To achieve proper coupling of the incident radiation with the cell objects Deng at al. (2019) proposed use of sub-wavelength apertures along with surface modification to allow high EMR transmission and generation of high extraordinary electric field concentration. For tight confinement of EMR they suggest a coupling mechanism that involves generation and propagation of surface plasmon polaritons (SPP) along a metal-dielectric interface and work with graphene waveguide with a nanoscale plasmonic resonator. They designed a sub-wavelength structure whose frequency is easily selectable by changing the rotational direction of the sub wavelength structure (a spiral bulls-eye structure). The system was tested for the frequency range 1-1.4 THz. Authors provide details of applying such frequency tunable SPP for multi frequency THz analysis. They explored applications in medical examination (local transmission spectra of pharmaceutical tablets and organ tissues). Biomolecule detection in THz domain is an important challenge especially because the THz vibration spectroscopy yields very useful results.

An overview of use of sub-millimeter waves for biology and medicine was published (Siegel, 2004) that pointed out the use of 300GHz-3THz EMR applications to study absorptive loss of biomaterials (linear attenuation coefficients of each type of tissue), in identifying the protein states and molecular signatures, tissue identification, disease detection, and THz microscopy. Another review done by authors Wei et al. (2019) points out that use of THz-TDS systems working in the 2-5 THz band can be utilized to study extensively the nucleic acids, protein, amino acids, and peptides, carbohydrates, and biological safety issues. Authors have pointed out that water absorption in these bands are a major problem and suggested use of microfluidic and microchannel devices to minimize water loss in signals. They have also suggested use of attenuated total reflection (ATR) technique, metamaterials (periodic artificial electromagnetic media smaller than the stimuli that enhances the local E field) and use of waveguides in experiments.

A not so recent review (Smye et al. 2001) in the UK have pointed out that since the typical photon energies associated with the THz in the above frequency domain are between $2 \times 10^{-22}$ Joules to $1.3 \times 10^{-20}$ Joules, interactions with bio-tissues would require deeper understanding of the "dielectric properties" of the biological materials such as permittivity and conductance of the medium that describe the bulk charge dynamics, and molecular dynamics once stimulated by the high frequency EMR. Molecular energy level transition study also becomes very important at these frequencies. They also point out, at frequencies below 6 THz, the interaction of EMR with bio sample behave classically, and can be well understood just by using parameters of permittivity and conductivity. However, at higher frequencies transitions between different molecular rotational and vibrational energy levels takes place, and the formalism of treating with a relevant theory would demand use of quantum mechanical framework. Authors have pointed out that more experimental data is required, and the higher frequency domain (6 THz-20 THz) should enable understanding the vibrational modes of the DNA.

In a recent work, Greschner et al. (2019) demonstrated that EMR in microwave and THz regimes affects the conformation of DNA in solution. They mention that THz radiation resonates with LF vibrational modes such as intra/inter-molecular interactions of molecules. In the 2.4-6 THz band they expect that the hydrogen-bond network of water is excited and stretching modes of double-stranded DNA (dsDNA) occur at 20 GHz, 500 GHz, 1 THz and at 6 THz. They also refer to the computational studies published that predict dsDNA possessing the high-density vibrational modes between 100 GHz to 6THz domain. In their experimental confirmation work they used picosecond pulsed THz (with repetition rate 2500 Hz) in the range 100 GHz to 3 THz (average power ~1 mW) and concluded that the EMR in microwave regimes promote the repair and assembly of DNA nanostructures while, intense THz pulses induce very rapid dissociation of short dsDNA in ambient environment.

Pickwell and Wallace (2006) published one review of THz technology in the 0.1-10 THz domain that was in contemporary use for biomedical applications. They have identified the different THz sources used for medical applications, outlined the terahertz pulsed spectroscopy (TPS) and imaging (TPI) techniques for biomedical applications citing examples of TPS for water, animal fats, protein, optical properties of human tissue, in vivo and in vitro imaging to reveal contrast between healthy skin and basal cell carcinoma (BCC), TPI for cortical bone, and crystalline molecules. Authors presented examples of TPI imaging techniques for skin, breast cancer, and teeth.

Terahertz EMR studies involve treating the stimulation waves partly from applying microwave electronic principles blended with photonics framework, since the EMR at these frequency range exhibit photon-like and electron-like (dual) properties as well.

Slides of an invited lecture at a Summer School on Optics & Photonics held in Oulu, Finland (Khodzitsky 2017) presented materials highlighting importance, and understanding the current state of the science in THz biology, in terms of studying THz-biomaterial interactions. Author points out that THz waves are unique, as it bears low photon energies, extreme water absorption, exhibit non-Mie scattering (bearing a longer wavelength than mid-infrared) resulting in transparency, strong molecular absorption due to dipole-allowed rotational and vibrational transitions prominent at THz frequencies. This study also points out that THz signals can be readily mapped for its transient changes of electric field, thereby suggesting strong candidacy of this technique for absorption and dispersion spectroscopy. Spectral signatures from different scales of motion (for THz frequencies) are explained in tabular form (Table 1) adapted from one of author's slides.

**Table 1**. The 3 different motion types detectable in biological materials, their respective time scales (based on Khodzitsky 2017) and frequency range for each motion type.

| Motion type | Time-scale | Frequency range |
| --- | --- | --- |
| Molecular Vibration | Sub-picosecond | NIR (200 THz-500 THz), FIR (3THz – 100 THz) (Infra-red) |
| Collective motions | Picoseconds | 300 GHz - 20THz (Terahertz) |
| Protein folding | Nano seconds ~ milliseconds | 100 KHz – 300 GHz (Microwave) |

Small biomolecules such as mononucleotides, fatty acids, monosaccharides, and amino acids exhibit clear spectral features due to less broadening and overlap effects. Low temperature measurements are ideal, as it minimizes spectral broadening. Transmission, reflection, and absorption spectroscopy is possible to be applied for small biomolecules. THz spectra for small organic molecules exhibit very clear features due to less broadening/overlap. Since water absorption is marked, small biomolecules need to be either dissolved in non-polar liquids or be of solid-state nature, and in general, low temperature measurements are ideal for small biomolecules because it minimizes spectral broadening. For macro biomolecules that have more monomers such as proteins, polysaccharides, nucleic acids (DNA, RNA), lipids exhibit strong spectral broadening. Phase change of the signal can be successfully utilized to evaluate aberration of the molecules. However, a far infra-red light may be used to agitate the vibrational and rotational energy of the large molecular system and the transients due to differential absorption of THz beam can be recorded for estimation of relaxation periods of the sample. However, for macromolecules, they do not yield clear spectral features in the THz bands, hence, a part of the beam gets transmitted, and a phase-change is introduced. This manifests itself to a changed refractive index and dielectric property of the large molecule, thereby yielding more details about the aberration. Proteins, for example, when excited by far infrared light (FIR), the excitation alters its vibration structure, and as a result will cause differential absorption of the passing THz probe beam through the sample. and we can study the decay of transient absorption in greater details to derive the relaxation periods of the large molecules. Thus, authors have pointed out that transmission, absorption, attenuated total reflectance (ATR) spectroscopy, and optical pump-THz probe spectroscopy are useful procedures. At biological cell level, different categories of cells, or same kinds of cells at different conditions may respond differently to THz waves hence, they are highly distinguishable and able to be characterized in detail. For tissues such as skin, bone, muscle, fat, liver, kidney, tooth, etc. (ensemble of similar cells) they are considered to have different water contents and hence show different responses to THz wave energy.

Some authors (Yaekashiwa et al. 2019) have also refuted presence of any non-thermal effects in the frequency range of 300 GHz-600GHz ( using weak power densities between $31\mu W/cm^2$ to $5mW/cm^2$ ) based on exposure studies of 2 sets of human cultured cell lines (cultured human skin fibroblasts and corneal epithelial). They used a 1 GHz resolution facile THz source for the 300-600 GHz band. They measured proliferative and metabolic activities before and after exposure and found no compelling evidence of change and no effect on the levels of HSP70 mRNA expression. Such attempts must be repeated and verified for results.

Sun et al. (2017) have published a review work highlighting use of THz systems (200 GHz-800 GHz) for in vivo and ex vivo biomedical studies with added review on understanding the THz imaging response datasets.

Distribution of the number of experimental works published in the entire 0.03 THz-20 THz domain is shown in Figure 6. Three important frequency bins have been identified.

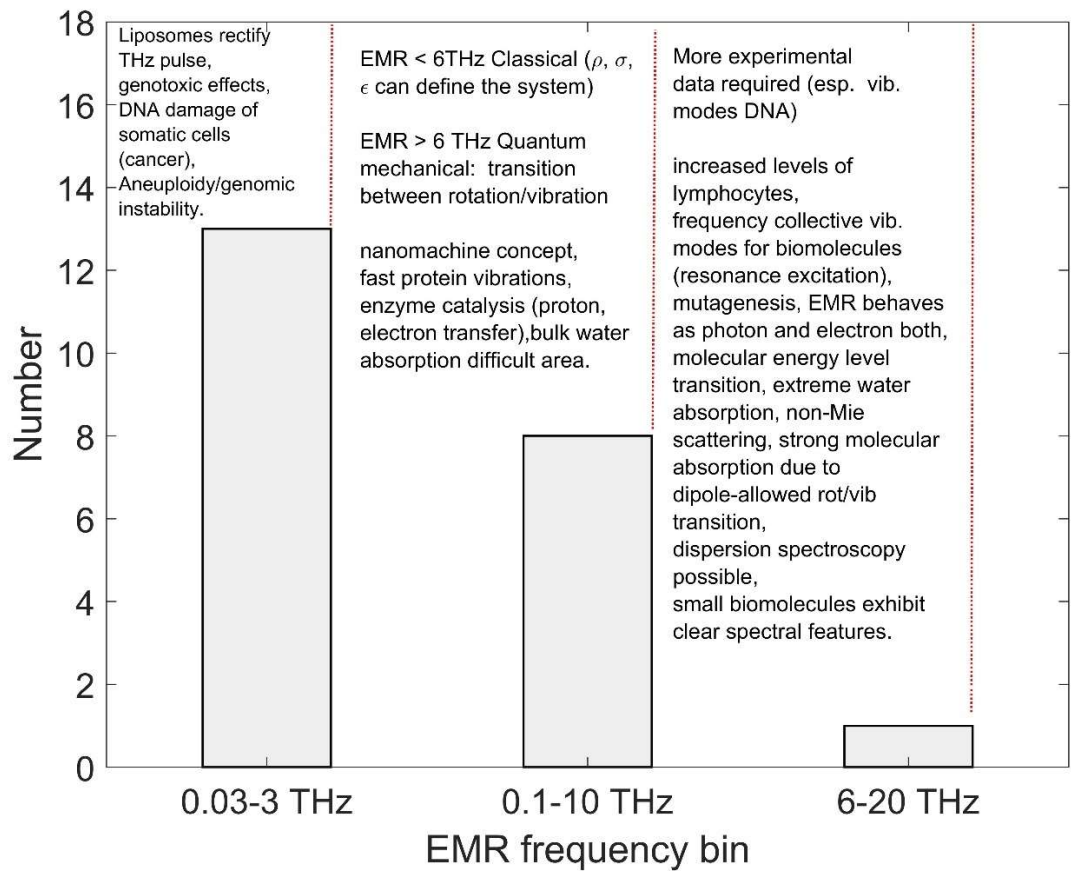

**Figure 6.** Bar plot showing the number of research work contributed to the 0.03-20 THz domain with dominant physical mechanisms, and significant bio-effects observed in each sub-band (bin).

We note that most of the published experimental reports are in the 30GHz-3THz range. 6-20THz range poses many significant physical challenges, and, combined with expense and general non availability of sources and detectors it has a smaller number of published experimental work.

**2.3 Biological effects in the FIR to near-visible band (20THz-461THz)**

This is a very wide band of optical radiation, including the far-IR, mid-IR, and the near-IR bands i.e., extending from THz to near visible region. The systems using these frequency ranges are very difficult to conceive, and sources are quite expensive. However, there is a lot of current and planned use of medical and industrial devices, that expose the human body to these optical and heat fields, hence, biological information about understanding the CW and pulsed exposure process, and their consequences are crucial for these EMR frequency ranges. A very useful dataset on optical property of biological tissues (with *in vivo* tissue parameters) for the 1THz-100 THz EMR has been published (Jacques, 2013).

In terms of biological impacts, research studies suggest that thermal to photochemical changes are mostly observed due to prolonged exposure with increase in frequency. Far infrared (FIR) systems deliver EMR having wavelengths in the region 1000 μm ~ 50 μm (corresponding to frequencies 300 GHz to 6 THz) energy ranging between 1.2 meV to 24.8 meV. The MIR systems deliver EMR with photons in energy levels in the range 24.8 meV to 0.413 eV (wavelengths 50 μm to 3.0 μm) in the frequency range 6 THz to 100 THz, and the NIR systems bearing wavelength in the range 3.0 μm to 0.78 μm (100 THz to 384.3 THz) with photon energies in the range 0.41 eV to 1.58 eV, respectively.

Biological effects revealed so far, in this wide EMR energy range have basis in absorption of energy by vibrational levels of bonds, in respective molecules. Vatansever and Hamblin (2012) point out that due to high concentration of water in biological systems, and association of water molecules with ions (solvation) the dielectric properties, and the large dipole moments results in strong interaction with EMR in the FIR range. Since tissue dielectric properties vary with water content, and show a strong dependence on the EMR frequency, the relaxation of these meso-structures can also show a frequency dependency. Local changes in the molecular environment (due to solvation, or confinement) affect the translational and vibrational modes in the FIR frequency range.

A ceramic FIR generator 25μm -3 μm (3 THz – 99 THz respectively) was used for irradiating human umbilical vein endothelial cells (HUVECs) (Hsu et al. 2012). Exposure power density around 130μWatt/cm$^2$ (non-thermal) for 30 min was accomplished. They summarized the findings that FIR exposure significantly inhibits the vascular endothelial growth factor (VEGF)-induced proliferation in HUVECs. They tried lower, and higher power densities, and found that the inhibition in proliferation rate is displayed consistently. They also report that FIR exposure significantly inhibited the phosphorylation of primary antibody ERK1/2 that peaked at 30 min exposure and gradually decreased at longer exposure periods and increased the phosphorylation of endothelial nitric oxide (NO) synthetase (eNOS). They attribute these changes as 'non-thermal' effect of FIR irradiation on HUVEC.

Ishibashi, et al. (2008) performed an *in vitro* set of experiments with FIR exposure technique in a tissue culture radiant panel incubator (coated with highly efficient ceramic material and employing polycarbonate printing technique) to continuously irradiate 5 different types of human cancer cell lines namely, A431 (vulva), HSC3 (tongue), Sa3(gingiva), A549 (lung) and MCF7 (breast). EMR in the frequency range 14.9 THz to 74.9 THz were used (with peaks at 24.9, and 42.8 THz). The FIR frequency range is at the body-temperature range. The incubator was maintained with 100% humidity at $37\pm0.5^0$ C and 5% of $CO_2$ in air. They calculated the total FIR energy emitted into the incubator was 161.283 Watts/steradian. Assuming isotropic emission, the total power was computed to be 2026.75 Watts, and considering the volume of the incubator 0.126 m$^3$ and volume of culture medium 6 ml amount of energy absorbed by each 10-cm culture dish was 9.67 mW. Based on the surface area of culture dish, the total energy absorbed was 4.43 J/hour cm.$^2$ With this kind of exposure they confirm that after exposure, the basal expression level of heat shock protein HSP70A mRNA was higher in A431 and MCF7 cells that in the FIR-sensitive HSC3, Sa3, and A 549 cells. They also report statistical data analysis on the cause of over expression of heat shock protein HSP70 that resulted in inhibited FIR-induced growth arrest in HSC3 cells, and an HSP70 siRNA (small interfering RNA) inhibited the proliferation of A431 cells after FIR irradiation. They conclude that, irradiating the cancer cell lines with FIR at the body temperature range suppresses proliferation of some cancer cells which are controlled by the basal expression level of the protein HSP70A, and hence, suggest FIR irradiation is suitable to treat cancer cells that have low levels of HSP70. The dosage and exposure can be adaptively adjusted if the level of HSP70 is measurable.

Results on the skin wound healing capability of FIR radiation were presented by few authors (Toyokawa et al. 2003). They investigated the heating effect of FIR radiation as well as the resulting biological effects on skin wound healing based on careful evaluation of the healing speed by comparing amongst groups, with and without FIR irradiation (using a rat model). Other parameters they documented for their analysis are the wound area, skin blood flow, skin temperature before, and after irradiation, and histological inspection. They report greater collagen regeneration and infiltration of fibroblasts, that expressed transforming growth factor beta1 in wounds that were exposed that without exposure.

Near infra-red (NIR) window is normally between 222 THz to 461 THz (wavelength, 1350 nm down to 650 nm). At this band light has maximum depth of penetration (DOP) in tissue and scattering dominates the EMR-tissue interactions. Photon absorption probability increases rapidly due to the occurrence of scattering. An excellent review of the tissue absorption profiles (by EMR frequency) of blood (at shorter wavelengths), water, melanin, and fat has been reported in literature (Jacques, 2013). This review paper also presents a set of formulae required for generating optical properties of a generic tissue with variable amounts of blood, water, melanin, fat, yellow pigments, and a variable balance between small-scale and large-scale scatterers present in the structures of cells and tissues.

Low level light therapy (LLLT) based photo-stimulation and photo-biomodulation (PBM) using the 3-394 THz EMR is being widely used for medical treatment (Tsai and Hamblin, 2018). Authors mention that the cellular and molecular action mechanisms of LLLT are now a days better understood and cites that most of the studies proposed that the chromophores responsible for PBM effects are primarily classified as mitochondrial photo-acceptor (cytochrome C oxidase, CCO). LLLT can increase the CCO enzyme activity to facilitate the electron transport hence, increasing the adenosine tri-phosphate production. This technique has clear medical benefits and for therapeutic delivery of IR is nowadays possible without using external power sources rather, using the body thermal energy and specialized materials to drive FIR energy.

## 3. Human safety and standards

Specific Absorption Rate (SAR) is the metric of the dose of EMR the human cell is exposed mostly for EMR frequency range 100 kHz-6 GHz range. This unit is defined as the power absorbed per unit mass generally the unit is Watts/Kg. SAR value is defined for 1 gram or 10 gram of biological tissue in the shape of a cube. SAR depends on the conductivity ($\sigma$) in unit of Siemens/cm, the electric field strength (E) in volts/meter, and the tissue density (g/cm$^3$). Neelakanta and Sharma (2013) have defined SAR very clearly in terms of the conductance $\sigma(r)$, electric field intensity $E(r)$, and density $D(r)$ of the bio medium as function of spatial variable (r)

$$SAR = \int \frac{\sigma(r)|E(r)|^2}{D(r)} dr \qquad (12)$$

For all practical purposes, the bulk expression for SAR = $\sigma|E|^2/\rho$ is utilized. It can also be related to the thermal response directly as SAR = C $\Delta T/\Delta t$ hence, SAR can be either determined from the electrically measured dielectric parameters or, from the thermal signature (temperature and exposure time). C in the thermal relationship is the specific heat capacity of the tissue. Exposure standards of SAR is laid down by the World Health Organization (WHO) and, for operational use and regulatory activities by the International Commission of Non-ionizing Radiation Protection (ICNIRP) and the United States it is practically regulated by the Federal Communications Commission (FCC). The safe accepted limit for SAR for mobile handsets is accepted to be 1.6 W/Kg (averaged over a cubical volume of 1 gram of tissue). SAR greater than 120 W/Kg becomes very harmful for biological tissues and become fatal for the person receiving that kind of radiation. Long term tissue heating might have carcinogenic effects. An analogous metric for pulsed power is current density (J) for the 100KH-10 MHz range (see Table 3).

Human exposures at short distance radiation due to indoor wireless connectivity (Wi-Fi) generally in microwave EMR frequencies (2.4, or 7 GHz) causes oxidative stress, apoptosis, cellular DNA damage, endocrine changes, and calcium overload (Pall 2018). Polarized and pulsed EMR is more active inactivation of voltage-gated calcium channel (VGCC) and that EMR exposure impacts are more prominent in young people than adults.

Cell phone EMR studies were conducted by the National Institute of Environmental Health Sciences (Wyde et al. 2018) using young, and aged rats and mice, and in pregnant rats to determine the effects of animal size and pregnancy status on the thermal response of EMR. Specific absorption rates (SAR) between 10-12 Watts/kg using 0.9 GHz radiation for approximately 9 hours per day for 5 consecutive days induce an increase in body temperature resulting in mortality of aged rats. At exposures of 8 Watts/kg, only pregnant rats showed a significant increase in body temperature.

Bhat and Kumar (2013) have shown that the human head can absorb the EMR radiation more readily, and about $0.3^0$ C change at the surface of the brain is observed for a hand receiver set operated at 0.9 GHz frequency for a sustained period of about 20 minutes. The heating patterns induced by the incident EMR on the biosystem will be dependent on the dielectric properties of the tissues (Johnson 1972). DOP can be measured or modeled and physically this depth pertains to the thickness through which the photon is delivered inside the tissue, and at which the incident intensity falls off to a value of 1/e or reduced to about 36.7% of the incident intensity. The rest of the EMR energy is expected to be absorbed in the tissue and, if the absorption is high then automatically, the penetration depth is low. Tissues with high water content such as muscle, brain, internal organ, and skin tissue have high water content hence the depth is lower. Penetration depth involves knowledge of the tissue's dielectric properties such as magnetic permeability ($\mu$), the conductivity ($\sigma$) in Siemens/m, and the relative permittivity ($\varepsilon_r$). Inan (2005) has presented a very detailed experimental dataset showing $\sigma$, $\varepsilon_r$, and DOP as a function of frequency (1-100GHz) for dry and wet skin, muscle, brain, and fat.

Moradi et al. (2016) pointed out in this paper that the ICNIRP has also set a magnetic field exposure limit for industrial workers to 60 mTesla or 600 Gauss. Internal fields are set up in bio-tissues when exposed to oscillating magnetic fields the upper limit of the frequency of oscillation is set at 30 kHz or less. The lower end of the oscillating magnetic field is set at 300 Hz.

The FCC report (1999) presented a useful table showing the frequency-dependent FCC Maximum Permissible Exposure (MPE) levels. Table 2 shows the prescribed magnitudes of the electric field (E), magnetic field (H) power density (S), and averaging time prescribed for each frequency range in MHz for controlled and uncontrolled scenarios, respectively.

**Table 2**. The FCC limits (FCC 1999) for maximum permissible exposure (MPE). *indicates plane-wave equivalent power density.

| Frequency (MHz) | Electric Field Strength E Volts/m | Magnetic Field Strength (H) Amps/m | Power Density (S) mW/cm$^2$ | Averaging Time $|E|^2$, $|H|^2$ or S (minutes) |
|---|---|---|---|---|
| **Occupational/Controlled Exposure** | | | | |
| 0.3-3.0 | 614 | 1.63 | 100* | 6 |
| 3.0-30 | 1842/f | 4.89/f | 900/f$^2$ * | 6 |
| 30-300 | 61.4 | 0.163 | 1.0 | 6 |
| 300-1500 | -- | -- | f/300 | 6 |
| 1500-100,000 (100 GHz) | -- | -- | 5 | 6 |
| **General population/Uncontrolled Exposure** | | | | |
| 0.3-1.34 | 614 | 1.63 | 100* | 30 |
| 1.34-30 | 824/f | 2.19/f | 180/f$^2$ * | 30 |
| 30-300 | 27.5 | 0.073 | 0.2 | 30 |
| 300-1500 | -- | -- | f/1500 | 30 |
| 1500-100,000 (100 GHz) | -- | -- | 1.0 | 30 |

A detailed summary of questions and answers related to the biological effects and potential hazards of EMR is presented in the FCC document "OET Bulletin 56" published in 1999 (FCC 1999). This document points out that from radiocommunication point of view, two areas of body, the eye, and testes are particularly vulnerable to RF heating by the incident EMR due to reduced blood flow in these organs. Rabbits, when exposed to 100-200 mW/cm$^2$ power density EMR exhibit cataract and temporary sterility was also observed. Whole body absorption (due to resonance) is maximum when the EMR is between 0.08GHz and 0.1 GHz depending on the size, shape, and height of the human subject. The bulletin also points out that EMR related "non-thermal" effects are so far very poorly understood. The document points out that further research is required to find out links between these non-thermal effects and human health.

Brain monitoring utilizes neural interfaces. In one publication (Zhao et al. 2013) a comprehensive data on the thermal properties of the bio-tissues contained in the human head model are presented. The maximum power receptions (by the FCC SAR limits) using implanted antennae were computed for 3 different dipole antenna geometries and the results demonstrated that longer length implanted antennae with lower operating EMR frequencies and shallower implantable depths will maximize RF power reception before violating safety limits. For the mmW ranges (6 GHz-300 GHz) power density (defined in Table 3) is the basis of exposure and beyond 300 GHz with frequencies extending into the THz and IR (up to 500 THz) range time-integrated radiance (TIR) forms the basis for safety assessment (see Table 3).

Theoretical estimates for tissue THz exposure safety thresholds have been published (Saviz et al. 2013) using the microthermal approach. Using permissible power densities of THz EMR in the 100 GHz-10THz range they have constructed dielectric models for the skin and cornea of the eye. Since the wavelength of THz EMR is much larger than the size of the microscale tissue structures, authors have used a well-defined dielectric mixing theory as mentioned in Zhao et al. (2013) such as the Maxwell-Garnet mixing model (Shivola 2000). This is also referred to as effective media theory (Choy, 1999) or a homogenization approach. In this approach, they consider electric field localization and small thermal gradients that buildup at smaller scales resulting in unpredictable hotspots. This formalism they term as "inclusion of biomaterials (IOB)" for use in the microthermal absorption process. Using Laplace's equation for a lossy sphere approximation of IOB they related the electric field strength of the IOB (subscript b) with that of ambient

$$E_b = E_a \left(\frac{\sigma_a + j\omega\epsilon_a}{(2\sigma_a+\sigma_b)+(j\omega(2\epsilon_a+\epsilon_b))}\right) \quad (13)$$

$E_a$ is the peak value of ambient field strength and the E strength is uniform over the IOB sphere hence absorbed power they derive using $P=0.5V_b\, E_b^2$ where $V_b$ is the volume of IOB sphere

$$P = \frac{E_a^2}{2} \frac{9\sigma_b(\sigma_a^2\omega^2\epsilon_a^2)}{(2\sigma_a+\sigma_b)^2+\omega^2(2\epsilon_a+\epsilon_b)^2} V_b \quad (14)$$

The imaginary part of complex permittivity ε" in conductivity σ = ωε", also includes the ionic loss. The electrical conditions for which the absorption will be maximum in condition $\varepsilon_b \to \varepsilon_0$ (free space permittivity). From (14) it is also possible to estimate the maximum value of the IOB conductivity ($\sigma_b$')

$$\sigma_b' = \sqrt{4\sigma^2 + \omega^2(2\varepsilon_a + \varepsilon_0)^2} \quad (15)$$

Using (14) and (15) authors have rederived the maximum power available ($P_a$) on IOB

$$P_a = V_b \frac{E_a}{2}\left[\frac{9\sigma_b'(\sigma_a^2+\omega^2\epsilon_a^2)}{(2\sigma_a+\sigma_b')^2+\omega^2(2\epsilon_a+\epsilon_0)^2}\right] \quad (16)$$

For low end of EMR frequencies they found predicted incident power density thresholds bear a good agreement with the current safety standards, however, for the higher frequencies, only the worst-case analysis provided new information that entails lower radiation levels than the current standards. Non-ionized plasma oscillations in the structure might also be a problem for these EMR frequencies. Hence, isometric oscillation response may be accounted as given by Eq. (6) of Jia et al. (2019).

SAR is important measure as it is used heavily for assessing human exposure in the crowded RF communication field for RF frequencies < 6 GHz. However, with the shift of frequencies to the higher ends of 40-60 GHz for the 5G and 6G communication bands SAR and power density methods should be combined in some way to establish limits of exposure. A new quantification of the rate of long-duration RF energy deposition technique for SAR determination has been proposed by refining the approach and including heat-diffusion and conduction effects in 3 dimensions (Alon et al. 2017). Normally, only Joule's heating and dielectric heating mechanisms are only considered for thermal estimates of SAR. But the authors argue that for accurate measurements of temperature rise the equipment radiator elements and electronic elements and the phantom conductance itself bear different time scales which should be accounted for to reduce uncertainties in short-scale temperature measurement. The current operational techniques neglect the heat diffusion in the biomaterials hence, a longer exposure period is required for estimating the thermal impact parameters. They introduce a magnetic resonance mapping technique to get a very high-resolution temperature difference map and consider generalized heat equation inversion framework (HEI) which accounts for heat dissipation due to proper boundary conditions, and diffusion. They instead, compute the point-wise SAR in a finite difference framework and then average SAR over the spherical volume of 1 and 10g (as defined by ICNIRP) around the area of interest. The technique yields good stability in temperature estimates and, they used an experimental data set for EMR at 1.9 GHz to verify reconstructed SAR distributions.

The blood-brain barrier (BBB) is an important functional item of the brain. Many researchers have reported that with mobile phone RF exposure there occurs leakage in the BBB causing hyperthermia and heat stress. Pelletier de Gannes et al. (2017) have refuted BBB leakage in mice and rats when exposed to RF at 0.9 GHz and 2.45 GHz (typical global system for mobile, GSM, and universal mobile telephone system (3G) frequencies). They exposed the animals at these RF frequencies for a period of 45 minutes to 4 hours and repeated exposures for up to 90 minutes per day for up to 104 weeks.

SAR ranged between 0.25 W/Kg to 4W/Kg and the brain averaged SAR (BASAR) ranged between 0.3-6 Watts/Kg. They concluded that there was no evidence of induced BBB leakage at these low-level exposures from the mobile handsets.

The SAR and power density prescribed limits for use in cellular communication in the USA, Canada, China, Japan, and Korea have been thoroughly inter-compared (Mazar 2016). In the USA, the Republic of Korea, and Canada the limit of exposure is set at 1.6 W/kg in 1g. At the 0.4-1.5 GHz cellular transmission including UHF TV bands the maximum allowed power density as prescribed by ICNIRP, Europe and the Republic of Korea is limited to 200 Watts/m$^2$ whereas, at the 0.3GHz-1.5 GHz range, the US and Japanese threshold are 150 Watts/m$^2$ (33% higher than ICNIRP 1998 threshold).

## 4. Overview

Based on the available literature we have pointed out that EMR exposure studies in the entire THz to NIR domain exposing the fact that dielectric properties (such as conductance, permittivity) of tissues become important in this frequency range. Molecular energy level transitions also become important. An overview of human exposure, and thermal bio-optical, and photo-biomodulation aspects applicable to the therapeutic use of EMR in the FIR to NIR bands. Some details about references used for human exposure levels and dose limits and how they are applicable for the RF/microwave, mmW, THz, and IR bands are cited. A short discussion is made about the usefulness of the depth of penetration of EMR in tissues concerning the exposure limits. Further, a note on human safety limits of exposures (SAR and power density limits) based on the FCC, WHO, IARC, ICNIRP, and other agencies, and relevant publications have been discussed. Low- and high microwave power densities in the 3-30 GHz range are discussed in the light of *in vivo* and *in vitro* studies. It is pointed out that microwave radiation deposits energy in cells and tissues and there are active non-thermal effects that exert torque on magnetic dipoles in bio-tissues resulting in membrane shock based on Fröhlich's hypothesis (Fröhlich 1968). A simulation study in the region 40-100 GHz has also been discussed. However, it is observed from the publication volume in the past 5-7 years that the bioeffect studies beyond 2 THz are rare, and there is a large gap between 2-30 THz that needs to be filled more rapidly to understand the consequences of low power density EMR due to planned radio communications on human health.

A broad description of the exposure regulatory standards has been presented in terms of the specific absorption rate (SAR) and EMR power density (in mW/cm$^2$). These data have relevance to researchers for experiment planning purposes so that biological impacts are studied with the purview of regulatory exposure limits and other constrained parameters to be able to focus strongly on assessing the gross human health effects due to such exposure

Further, we essentially collect information from a broad range of EMR frequencies (kHz to NIR-384 THz range) and attempt to tabulate the human exposure parameters with respective metrics and safety limits for broad application areas of medicine, studies related to cell level biological effects and biophysical interpretations. Figure 7 provides these significances by a frequency line, and Table 3 is the companion to this figure that explains the abbreviations and frequency range denoting alphabets used in the figure. Appropriate Journal references from which the information is collected are cited.

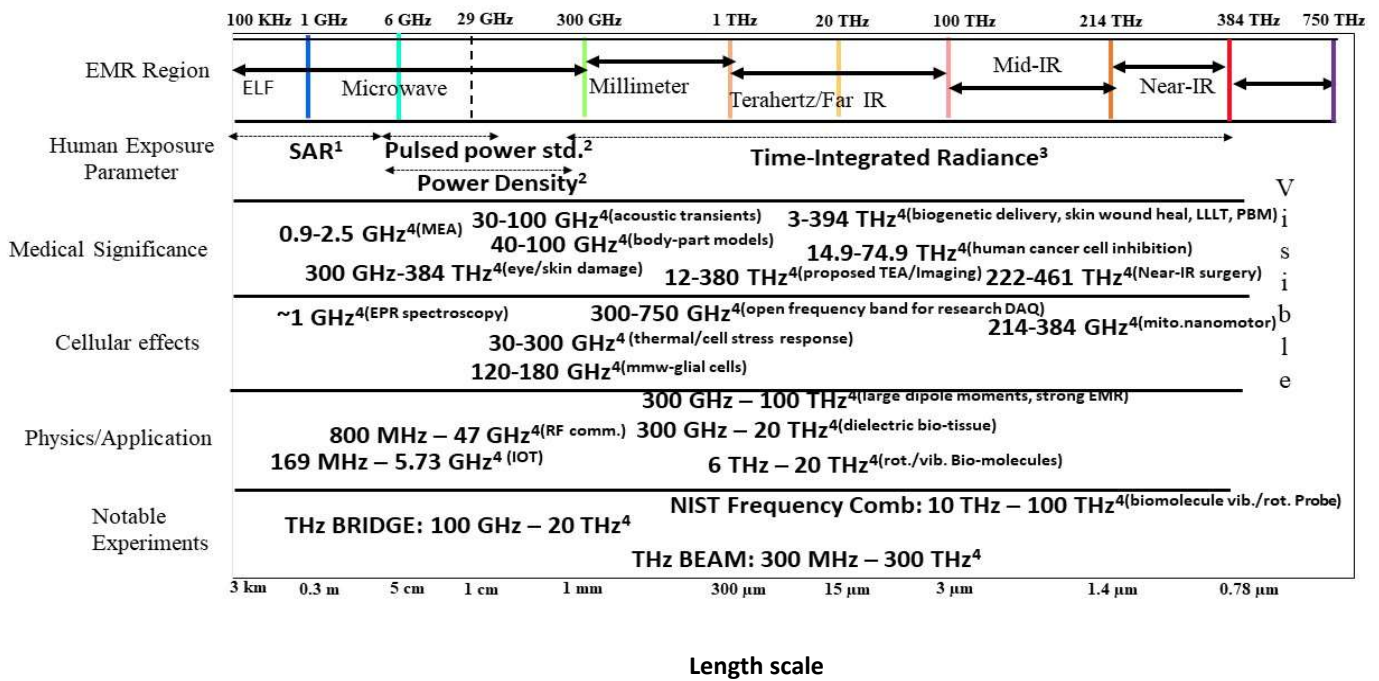

**Figure 7** EMR bands with human exposure parameters and standards (published, up to date), Frequency band's significance for medical treatments, cellular application domains and biological effects, the physical significance of EMR bio-responses at a molecular level, and some major experimental arrangements performed and planned for current & future. The lettered frequency markers a-e, and A-S are explained (with appropriate references) in Table 3 below.

**Table 3:** Superscript letters (a through e) and the frequency-range alphabet A-S used in the frequency summary (Figure 2) are explained in this table with exact frequency range in kHz/GHz/THz, scientific significance of the frequency range, and appropriate references in peer-reviewed journals.

| Superscript letter/Frequency range Alphabet used in Figure 2 | Frequency Range (GHz/THz) | Explanation/Scientific Significance | Reference |
|---|---|---|---|
| a, current density (J) specific absorption rate (SAR) standard | J: 100 kHz – 10 MHz<br><br>SAR: 100 kHz – 6 GHz | Current density (J)=σE, SAR is Ratio of EMR power loss density (W m$^{-3}$) to density of human tissue (Kg m$^{-3}$) expressed in terms of electric field E in V m$^{-1}$, $\frac{\sigma_i |E|^2}{2\rho_i}$ for the ith tissue | Safe range: 1.6 – 120 W Kg$^{-1}$, ICNIRP (2018), Moradi et al. (2016), Zhadobov et al. (2011) |
| b, Power density (PD)/Pulsed Power Density (PPD) standard | PD: 6 GHz -300 GHz<br><br>PPD: 6 GHz – 100 GHz | PD is ratio of incident EMR power to the exposed surface area expressed in mW cm$^{-2}$, PD=|E X H|, H is magnetic field in Amps cm$^{-1}$ | Safe range averaged over 20 cm$^2$: 1 mW cm$^{-2}$ (public) in 60 GHz band, ICNIRP/IEEE, Health Physics (2013), Zhadobov et al. (2011) |
| c, Time-integrated radiance (D)/radiance dose standard | 300 GHz – 384.34 THz | PD also used, Ocular exposure measured using time-integrated radiance, L in W m$^{-2}$sr$^{-1}$, $D = \int L(t)dt$ | PD defined in Saviz et al. (2013); Retinal thermal exposures are tabulated in ICNIRP (2013) paper (table 4), Health Physics (2013) |
| A | 0.9 GHz – 2.5 GHz | MEA: microwave endometrial ablation. Usually employed in shallow ablation of endometrial linings during medical procedures | Neelakanta and Sharma (2013) |
| B | 30 GHz – 100 GHz | Acoustic transients observed due to pulsed EMR resulting in transient heating of water | Pattanaik (2012) |
| C | 40 GHz – 100 GHz | Body part models, 4 different models adopted for 4 body parts and EMR power reflection coefficients ascertained using a multi-layer approach | Wu et al. (2015) |
| D | 14.9 THz – 74.9 THz | Human cancer cell inhibition band (anti-tumor), therapeutic applications, low-level light therapy (LLLT), sportswear | Tsai & Hamblin (2018) |
| E | 12 THz – 380 THz | Terahertz endometrial ablation (TEA) proposed | Neelakanta & Sharma (2013) |
| F | 3 THz – 100 THz | Biogenetic Far IR delivery, skin wound healing, photo-biomodulation (PBM) | Tsai & Hamblin (2018) |
| G | 300 GHz- 384 THz | Eye/skin damage treatment | Tsai & Hamblin (2018) |
| H | 222 THz - 461 THz | Non-ionizing near infra-red (NIR)surgery | Tsai & Hamblin (2018) |
| I | ~ 1 GHz | Electron paramagnetic/spin resonance spectroscopy considers of the energy difference between lower and upper state of unpaired electron in tissue sample in presence of external H | Tsai & Hamblin (2018) |
| **J** | 120 GHz – 180 GHz | In rat-glial cells: after 1 minute of exposure, a relative number of apoptotic cells increased 1.5 times and after 3 minutes the number doubled | Borovkova et al.(2017) |
| **K** | 30 GHz – 300 GHz | Thermal/Non-thermal effects such as shocks/membrane excitation observed in tissues due to electric field on charge and magnetic fields create torque on tissue magnetic dipoles. Mammalian cellular stress response (CSR) | Wilmink et al. (2006, 2007, 2008, 2009, 2011) |
| **L** | 300 GHz – 750 GHz | Very few data set available – an open area of investigations of EMR with biological structures and related bioeffects due to difficulties in cost, unavailability of reliable EMR source | |
| **M** | 214 THz-384 THz | Non-ionizing IR radiation can deliver small amount of vibrational energy to nanostructured water layers – this is established "mitochondrial nanomotor" technique | Sommer et al. (2008) and mentioned in Tsai & Hamblin (2018) |
| **N** | 0.3 THz – 331 THz | Anti-tumor action | Tsai & Hamblin (2018) |
| **O** | 160 THz-384 THz | Neural Stimulation | Tsai & Hamblin (2018) |
| **P** | 169 MHz-5.73 GHz | Internet of Things (IOT) transmission frequencies, will be widely used in medical facilities | R. Islam et al. (2015) |
| **Q** | 800 MHz-47 GHz | Radiofrequency (RF) communication including high 5G (fifth generation technology standards for wireless cellular communication) | https://www.fcc.gov/5G |
| **R** | 6 THz-20THz | Rotation/vibration of biomolecules, molecular energy level transitions, Quantum framework achievable, non-Mie scattering, powerful mapping capability due to dipole allowed transitions in transient changes of E field. | Smye et al. (2001) |
| **d, THz -BRIDGE, 2001-2004** | 100 GHz – 20 THz | Project generated spectroscopic database for enzymes, bio-membranes and cells using controlled experiments, assess critical EMR frequencies that can damage the biomaterial and to define safe exposure standards for THz imaging techniques | www.frascati.enea.it/THz-BRIDGE, Wilmink et al. (2011) |
| **e, THz BEAM, ongoing** | 300 MHz – 300 THz | • Develop THz sources & components<br>• Spectroscopic techniques<br>• Biomaterial interaction of THz beam | http://webusers.fis.uniroma3.it/sils/infrastutture/<br><br>grandi_infrastr_THz-BEAMrev.pdf, Wilmink et al. (2011) |

| | | • Sensing, imaging, and advanced materials | |
|---|---|---|---|
| f, NIST (National Institute of Standards and Technology) frequency comb | 60 GHz – 100 THz | • Disease diagnosis<br>• Fingerprinting antibody reference materials for biopharma.<br>• Can detect E fields of absorbed light in biomaterials | https://www.nist.gov/news-events/news/2019/06/nist-infrared-frequency-comb-measures-biological-signatures<br><br>Kowligy et al. (2019) |

Besides, we came across research materials that describe long-term exposure to mobile communication frequency on mice. This has shown to have enhanced amino acid neurotransmitters resulting in anxiety in subjects. Pulsed radiation shorter than 10 s in the 6-100 GHz range is sensitive to a more detailed derivation of actual energy absorbed per unit area and some authors have performed detailed studies using finite difference time domain (FDTD) and suggested inclusion of this factor for safety limit settings, which are currently dependent on EMR power density only. A simulation study for different body parts exposed to EMR at 40-100 GHz has shown that about 34% to 42% power is reflected off the air/skin interface due to variation in skin dielectric parameters and hence, suggested use of multilayer skin models for accurate results. A direct relationship between the EMR exposure time for rat glial cells and the relative number of apoptotic cells was noted by authors and they attribute this to direct resonance absorption in the 120-180 GHz range arising to the cell sizes being small compared to the wavelengths of exposure and that the glial cells are polarized. Ambient humidity plays an important role in heat deposited in cells due to EMR exposure in the microwave band especially in ocular cells.

## 5. Conclusions

We present a broad review of published data on thermal and non-thermal interactions of a biological cell, and tissue exposure to low power density nonionizing EMR extending from ELF to NIR ranges, and plausible explanations of the effects happening at the cellular level. In the beginning, we have pictorially explained the entire EMR spectrum to show the length-scale, energy- and temperature scale of the mid-end frequency domain that is poised for use by RF communication devices soon. The introduction section describes the components of EMR interactions such as due to RF communications (mostly thermal effects on cells), non-ionizing higher frequency EMR descriptions of non-thermal effects, and some aspects of IR therapeutic use of spectrum. Two big Italian experimental projects THZ BRIDGE and THZ-BEAM are described in this context. Some biological experiments and outcomes are also discussed briefly in this section.

We mentioned the outcome of a recent 10.5 GHz exposure studies for prostate cancer (PC) cells from 4 different cell lines and that the PC cells exhibited expression of phosphorylation of serine 2448 in RWPE2 and MDA-PCa2B and phosphorylation of serine 473 of AKT kinase in LNCaP and PC3 cells.

A discussion is made about the extent of EMR reflection from air-skin interface suggesting that 26-41% of incident EMR is reflected off this interface and clothing enhances transmission. We also discussed pulsed microwave EMR and transient heating of the water and subsequent expansion and mentioned work that supports the fact that non-thermal mechanisms occur mostly due to forces on charges due to electric fields and torques due to magnetic fields.

A concept of "evolved molecules" involving EMR frequencies in the range of 100 GHz to 10 THz and investigating the dynamical properties in a very wide time range (from sub-ps to ultralow periods) is explained. The scattering properties of the THz biomaterial interactions are discussed. We also discussed active studies of transmission, absorption, dielectric attenuated total reflectance, and differential spectroscopy techniques. Skin absorption of EMR is described in detail.

For the 300 GHz–20 THz band, generally, most of the authors have pointed out that for THz frequency ranges, frequency-dependent dielectric properties of the biomaterial become very important to study the outcome and effects. This is a domain where there are experimental difficulties along with the level of understanding at the cellular level interactions. Most of the authors have pointed out that the availability of THz sources, relevant optics, and detection systems are very expensive, and are hard to obtain in general. Authors have also pointed out that for EMR frequencies < 6 THz classical treatment of the interaction problem can be dealt with easily with the dielectric treatment but beyond 6 THz the interaction and outcomes are mainly controlled by the transitions between different molecular rotational and vibrational translations. Proper ascertainment of multiple reflections of EMR becomes quite important in treatments studying THz exposure behaviors at the cell/tissue levels. Some authors have proposed treating the subject of utilizing the "nanomachine" concept of the biomolecules at these frequency levels and have, in general, expressed treatment of bulk water in cells become very difficult especially biomolecule-water coupling mechanisms. At these frequencies, scattering effects are less, as a result, mapping of transient THz signals (time-resolved) may be possible, hence, some authors suggested the use of the principles of dispersion spectroscopy. Few authors have suggested the use of phase change in THz biomolecular signal to understand aberrations induced, and they also recommend the use of transmission, dielectric ATR, and differential techniques for studying different categories of cells. For tissues, which generally have different water content, transmission, reflection, absorption, dielectric, and ATR spectroscopy are recommended. Authors suggest that for human skin, long term EMR exposure at THz frequencies, skin components such as melanin and collagen structure in the dermis must be accounted for when estimating quantitative absorption.

Far infrared (FIR) and near-infrared (NIR) interactions are characterized by strong Rayleigh and Mie class scattering of EMR at the cell level. Not many experimental papers are published in this area but, theoretical modeling papers have appeared in a large volume. As a result, we do not have too many experimental data showing biological cell and tissue level responses at these frequencies. Some authors have shown that local changes in the biomolecular environment (due to solvation) affect the translational and vibrational models at the FIR frequency range. A systemic exposure study revealed that FIR exposure of vascular endothelial growth factor is inhibited attributing these to non-thermal effects (possibly dielectrophoretic forces) of FIR interaction with biomaterials. Another interesting work performed in a radiant $CO_2$ chamber suggests that FIR irradiation of cancer cells with low content of heat shock protein (HSP 70A) may be beneficial for cancer treatment. At NIR frequencies scattering dominates resulting in a higher probability of photon absorption, and EMR exhibits a very high depth of penetration in tissues. One recent theoretical modeling reference is provided that describes estimating the optical properties (extinction) at NIR frequencies for blood, water, melanin, fat, and yellow pigments. Low-level light therapy (LLLT) for therapeutic use and some functionality have been explained based on a recent publication.

A new direction to explain non-thermal biological effects manifests from the use of a proposed bio-soliton (solitary waves) theory (Geesink and Meijer, 2017). Such a mechanism is based on electron/phonon coupling of coherent standing waves based on 12 identified coupled oscillators. Broadly, this approach identifies "acoustic wave functions" associated with biological radiation data in Hz to THz/PHz (peta-Hertz) range and locates Eigenfrequencies of such a function. The model intends to support the concept of coherent quantized electromagnetic state in living organisms following in the line of studies done on materials.

With the current state of the science in exposure hazards, recorded at the cellular and tissue levels, due to absorption of EMR at a wide range of frequencies we strongly propose that full-scale laboratory experiments must be pursued to have a more solid understanding of the RF-biomaterial interactions, and deeper understanding of the gross human health effects of such radiations for prolonged periods especially in the light of projected future higher (up to 1.2 THz ) possible telephone networks that are being planned, medical diagnostics that use nonionizing radiation sources and low-level light therapies that are currently in an experimental phase. With the advent of techniques of deep learning spectroscopy employing artificial intelligence methods (Ghosh et al. 2019), direct training of the network can be performed based on the molecular structure of the investigative sample, and peaks with high accuracies can then be compared with experimental evidence at those EMR frequencies obtained by laboratory spectroscopy methods. Adoption of such methods may prove to be beneficial.

**Acknowledgments**

BR acknowledges the University of North Carolina Research Opportunities Initiative (ROI) through the NC Carbon Materials Initiative and the Center for Hybrid Materials Enabled Electronics (CH-MEET). BR was also partially supported by Dr. B. Vlahovic of NCCU CREST center and through the U.S. Department of Homeland Security grant 2016-ST-062-00004. We also gratefully acknowledge the grants 1U01CA194730, 1U54MD012392, 1R01MD012767 from the National Institutes of Health to BBRI.

**Declaration of Interest**

All authors of this paper declare that none of them have any conflict of interest in this work presented.

**Biographical Note**

Biswadev Roy is a research faculty at the Department of Mathematics & Physics, North Carolina Central University, Durham, North Carolina, U.S.A. He is engaged in experimental methods of studying materials using the microwave, millimeter, and THz waves as probe signals. He is interested in studying microwave/millimeter-wave properties in the condensed media, biological samples, and atmosphere using quasi-optical techniques.

Suryakant Niture is a research scientist at the Julius L. Chambers Biomedical/Biotechnology Research Institute (BBRI) at North Carolina Central University at Durham, North Carolina, U.S.A., and is engaged in cancer biology research.

Marvin H. Wu is a professor of Physics in the Department of Mathematics & Physics at North Carolina Central University, Durham, North Carolina, U.S.A. He is a condensed matter (solid-state) experimental physicist engaged in optical and microwave/millimeter-wave characterization of materials and currently the principal investigator of the millimeter-wave characterization instrument developed by him and BR.

**Data availability statement**

Data used in this study are published by each author cited in the reference list most of them have DOI numbers assigned and some of them also possess the URL link in the reference itself. Data of the 10.5 GHz irradiation of prostate cancer cells and further studies of their fate are available from the corresponding author upon request.